\documentclass[pdflatex,sn-mathphys-num]{sn-jnl}% Math and Physical Sciences Numbered Reference Style
\usepackage{graphicx}%
\usepackage{amsmath,amssymb,amsfonts}%
\usepackage{amsthm}%
\usepackage{mathrsfs}%
\usepackage[title]{appendix}%
\usepackage{xcolor}%
\usepackage{textcomp}%
\usepackage{manyfoot}%
\usepackage{booktabs}%
\usepackage{algorithm}%
\usepackage{algpseudocode}%
\usepackage{listings}%

\usepackage{graphicx}
\usepackage{xurl} % 专门处理url断行
\usepackage{seqsplit} % 如果还想自动拆分长词
\usepackage{amssymb}
\usepackage{multirow}
\usepackage{array}
\usepackage{amsmath}
\usepackage{mathtools}
\usepackage{threeparttable}
\usepackage{bm}
\usepackage{xcolor}
\usepackage{url}
\usepackage{booktabs}
\usepackage{subcaption} % subtable 环境
\usepackage{verbatim}
\usepackage{marvosym}
\usepackage{makecell}
\usepackage{tikz}
\usetikzlibrary{calc}
\usetikzlibrary{crypto.symbols}
\usetikzlibrary {positioning}
\usetikzlibrary{shapes}
\usepackage{diagbox}
\usepackage{adjustbox}
\usepackage{float}
\usepackage{longtable}
\usepackage{tabularray}
\UseTblrLibrary{booktabs} 
\usepackage{xspace}

\newcommand{\cipher}{\textsf{Gleeok}\xspace}
\newcommand{\cipherone}{\textsf{Gleeok-128}\xspace}
\newcommand{\ciphertwo}{\textsf{Gleeok-256}\xspace}
\newcommand{\branchone}{\textsf{Branch1}\xspace}
\newcommand{\branchtwo}{\textsf{Branch2}\xspace}
\newcommand{\branchthree}{\textsf{Branch3}\xspace}
\theoremstyle{thmstyleone}%
\newtheorem{theorem}{Theorem}%  meant for continuous numbers
\theoremstyle{thmstyletwo}%
\newtheorem{example}{Example}%

\theoremstyle{thmstylethree}%

\begin{document}

\title[Cryptanalysis of \cipherone]{Cryptanalysis of \cipherone}


\author[1,2]{\fnm{Siwei} \sur{Chen}}\email{chensiwei\_hubu@163.com}

\author[1,2]{\fnm{Peipei} \sur{ Xie}}\email{xiepeipei@stu.hubu.edu.cn}

\author*[2,3]{\fnm{Shengyuan} \sur{Xu}}\email{xushengyuan@sdust.edu.cn}

\author[4]{\fnm{Xiutao} \sur{Feng}}\email{fengxt@amss.ac.cn}

\author[1,2]{\fnm{Zejun} \sur{Xiang}}\email{xiangzejun@hubu.edu.cn}

\author[2,5]{\fnm{Xiangyong} \sur{Zeng}}\email{xzeng@hubu.edu.cn}

\affil[1]{School of Cyber Science and Technology, Hubei University, Wuhan, Hubei, China}

\affil[2]{Key Laboratory of Intelligent Sensing System and Security, Ministry of Education, Hubei University, Wuhan, Hubei, China}
\affil[3]{Department of Fundamental Courses, Shandong University of Science and Technology, Taian, Shandong, China}
\affil[4]{Key Laboratory of Mathematics Mechanization, Academy of Mathematics and Systems Science, Chinese Academy of Sciences, Beijing, China}

\affil[5]{Faculty of Mathematics and Statistics, Hubei Key Laboratory of Applied Mathematics, Hubei University, Wuhan, Hubei, China}


\abstract{\cipher is a family of low-latency keyed pseudorandom functions (PRF) including two variants called \cipherone and \ciphertwo, which are based on three parallel SPN-based keyed permutations whose outputs are XORed to produce the final output. Both \cipherone and \ciphertwo employ a 256-bit key, with block sizes of 128 and 256 bits, respectively. Due to this multi-branch structure, evaluating its security margin and mounting valid key-recovery attacks present non-trivial challenges. In this paper, we present the first comprehensive third-party cryptanalysis of \cipherone, including a two-stage MILP-based framework for constructing branch-wise and full-cipher differential-linear (DL) distinguishers, and a dedicated key-recovery framework based on integral distinguishers for multi-branch designs. Our analysis yields 7-/7-/8-/4-round DL distinguishers for \branchone/\branchtwo/\branchthree/\cipherone with squared correlations $2^{-88.12}$/$2^{-88.12}$/$2^{-38.73}$/$2^{-49.04}$. All distinguishers except the one targeting the PRF outperform the best distinguishers given in the design document. Moreover, by tightening algebraic degree bounds, we obtain 9-/9-/7-round integral distinguishers for the three branches and a 7-round distinguisher for the full PRF, extending the existing ones proposed in the original design document by 3/3/2 rounds and 2 rounds, respectively. Furthermore, the newly explored integral distinguishers enable the key-recovery attacks on \cipherone: a 7-round attack in the non-full-codebook setting and an 8-round attack in the full-codebook setting. In addition, we uncover a flaw in the original linear security evaluation of \branchthree, showing that it can be distinguished over all 12 rounds with data complexity $2^{48}$, and propose optimized linear-layer parameters that significantly strengthen its linear resistance without sacrificing diffusion. These results advance the understanding of \cipherone's security and provide generalized methods for analyzing multi-branch cipher designs.}


\maketitle

\section{Introduction}
In recent years, the growing demand for real-time applications, such as 5G communications, Trusted Execution Environments (TEEs), telemedicine, and vehicle-to-everything (V2X) systems, has made cryptographic latency a critical performance factor. In such latency-sensitive contexts, cryptographic operations often lie on the critical path, and even small delays can degrade system responsiveness. To address this, several low-latency symmetric designs have been proposed, including \textsf{PRINCE}~\cite{DBLP:conf/asiacrypt/BorghoffCGKKKLNPRRTY12}, \textsf{QARMA}~\cite{DBLP:journals/tosc/Avanzi17}, \textsf{MANTIS}~\cite{DBLP:conf/crypto/BeierleJKL0PSSS16}, and \textsf{Orthros}~\cite{DBLP:journals/tosc/BanikILMS21}. \textsf{Orthros}, for example, is a 128-bit keyed pseudorandom function (PRF) constructed as the sum of two keyed permutations to achieve security through structural composition. However, most existing designs support only 128-bit keys, which may be insufficient for emerging applications requiring higher security levels. Against this backdrop, \cipher was recently introduced as a family of low-latency PRFs supporting 256-bit keys while maintaining sub-nanosecond latency~\cite{DBLP:journals/tches/AnandBCIILMRS24}. 

\cipher is a recently proposed family of pseudorandom functions designed to achieve ultra-low latency while supporting 256-bit keys. Drawing inspiration from Orthros, its core construction is based on three parallel 128-bit branches: the first two are tailored to provide resistance against statistical attacks, particularly differential cryptanalysis, while the third branch is intended to counter algebraic attacks. The outputs of the three branches are XORed to form the final ciphertext. In addition to the standard variant, \cipher also includes wide-block versions using three 256-bit branches. Currently, there is no third-party analysis of \cipher, and analysis frameworks for multi-branch cipher structures remain limited.

The original design document of \cipher evaluates its security against various classes of classical attacks, including differential and linear cryptanalysis, impossible differential attacks, integral attacks, and other related methods. Their analysis concludes that the best known attacks on \cipherone and \ciphertwo are 5-round integral distinguishers rather than key-recovery attacks. Since decrypting the last rounds from the ciphertext would require guessing 256 bits for \cipherone or 512 bits for \ciphertwo, incorporating a key-recovery phase based on these distinguishers was declared to be an open problem. Furthermore, to date, no third-party cryptanalysis has been conducted to independently validate or challenge these security claims.

Motivated by these gaps, this work aims to revisit the security claims of \cipher. In particular, we aim to investigate whether longer and more effective distinguishers can be constructed, and whether practical key-recovery attacks are feasible despite its multi-branch structure. Answering these questions is essential for a comprehensive understanding of \cipher's security margin and for advancing cryptanalysis methodologies for similar designs.

\subsection{Our Contributions}
In this paper, we conduct the first comprehensive third-party cryptanalysis of \cipherone, targeting both its individual branches and the overall PRF structure. Our work integrates differential-linear (DL) and integral cryptanalysis as well as a novel key-recovery framework tailored for multi-branch designs. The main contributions are summarized as follows:
\begin{itemize}
    \item \textbf{Two-stage MILP-based framework for branch-wise DL distinguishers.}
We present a generic two-stage MILP-based framework for the automated search and precise evaluation of high-quality DL distinguishers. Using this framework, we identify a diverse set of efficient distinguishers for \cipherone: for \branchone\ and \branchtwo, we construct 7-round DL distinguishers with squared correlation $2^{-88.12}$, improving the data complexity over the previously best-known 7-round boomerang distinguishers; for \branchthree, we obtain an 8-round DL distinguisher with squared correlation $2^{-38.73}$; for the full \cipherone, we derive a 4-round distinguisher with squared correlation $2^{-49.04}$. These DL distinguishers are compared with the existing ones in Table~\ref{tab:result}. In addition, lower-round results are experimentally verified, with measured correlations closely matching the MILP-based estimates, confirming the accuracy of our framework.

\vspace{3pt}
\item \textbf{Algebraic-degree-based integral distinguishers and key-recovery framework.}
We apply an algebraic-degree-bound approach for constructing longer and more efficient integral distinguishers than those obtained via division property. This method yields 9-, 9-, and 7-round integral distinguishers for \branchone, \branchtwo, and \branchthree, respectively, and a 7-round distinguisher for \cipherone. Building on these results, we design a key-recovery framework tailored for multi-branch ciphers that avoids guessing intermediate states of all branches by prepending only one round before the distinguisher. Using this framework, we mount a 7-round key-recovery attack in the non-full-codebook setting with $2^{124}$ chosen plaintexts and time complexity $2^{133.6}$ (further reduced to $2^{132}$ with precomputation), and an 8-round attack in the full-codebook setting that recovers all 256 key bits with time complexity $2^{129}$ and memory $2^{133}$ bytes. These represent the first key-recovery attacks on \cipherone to date. The comparison of our newly explored distinguishers with the previous ones are summarized in Table~\ref{tab:result}, and details of our key-recovery attacks are shown in Table~\ref{tab:key-recovery}.

\vspace{3pt}
\item \textbf{Revised linear security evaluation and optimized linear-layer parameters.}
We identify a modeling error in the original security evaluation of \branchthree, where the linear mask propagation through the $\theta$ operation was incorrectly specified. Correcting this reveals that \branchthree\ can be distinguished over all 12 rounds with expected data complexity $2^{48}$, indicating significantly weaker linear resistance than originally claimed. To mitigate this weakness, we perform an exhaustive search over $\theta$/$\pi$ parameter sets that preserve full diffusion, identifying six improved parameter classes with markedly lower maximal squared correlations in the early rounds. Among these, the Improved Class B (see Table~\ref{tab:NewParams}) achieves the stronger resistance to linear cryptanalysis and differential cryptanalysis without compromising the original design rationale.
\end{itemize}

\begin{table}[!htbp]
    \centering
    \small
    \caption{The distinguishing attacks on \cipherone and its underlying branches. The full version has 12 rounds.}
    \label{tab:result}
    \setlength{\tabcolsep}{5pt}
    \begin{tabular}{ccllc}
    \toprule
     Cipher & Round & Type & Data & Ref.\\\midrule
     \multirow{5}{*}{\textsf{Branch1/2}} & 6 & Integral & $2^{127}$ &~\cite{DBLP:journals/tches/AnandBCIILMRS24}\\
     & \textbf{6} & Integral & \bm{$2^{65}}$ &Sect.~\ref{sect:integral_distinguisher}\\
     & 7 & Boomerang & $2^{100}$ &~\cite{DBLP:journals/tches/AnandBCIILMRS24}\\
     & \textbf{7} & Differential-linear & $\bm{2^{88.12}}$ &Sect.~\ref{sect:DL-distinguisher}\\
     & \textbf{9} & Integral & $\bm{2^{126}}$ &Sect.~\ref{sect:integral_distinguisher}\\\midrule
     \multirow{7}{*}{\textsf{Branch3}} & 5 & Integral & $2^{127}$ &~\cite{DBLP:journals/tches/AnandBCIILMRS24}\\
     & \textbf{5} & Integral & \bm{$2^{113}}$ &Sect.~\ref{sect:integral_distinguisher}\\
     & 8 & Differential & $2^{64}$ &~\cite{DBLP:journals/tches/AnandBCIILMRS24}\\
     & 8 & Linear & $2^{64}$ &~\cite{DBLP:journals/tches/AnandBCIILMRS24}\\
     & \textbf{8} & Differential-linear & $\bm{2^{38.73}}$ &Sect.~\ref{sect:DL-distinguisher}\\
     & \textbf{8} & Linear & $\bm{2^{32}}$ &Sect.~\ref{sect:linear_distinguisher}\\
     & \textbf{12} & Linear & $\bm{2^{48}}$ & Sect.~\ref{sect:linear_distinguisher}\\\midrule
     \multirow{6}{*}{\cipherone} & 4 & Differential & $\geq 2^{126}$&~\cite{DBLP:journals/tches/AnandBCIILMRS24}\\
     & 4 & Linear & $\geq 2^{124}$&~\cite{DBLP:journals/tches/AnandBCIILMRS24}\\
     & \textbf{4} & Differential-linear & $\bm{2^{49.04}}$ & Sect.~\ref{sect:DL-distinguisher}\\
     & 5 & Integral & $2^{127}$ &\cite{DBLP:journals/tches/AnandBCIILMRS24}\\
     & \textbf{5} & Integral & $\bm{2^{113}}$ &Sect.~\ref{sect:integral_distinguisher}\\
     & \textbf{7} & Integral & $\bm{2^{127}}$ &Sect.~\ref{sect:integral_distinguisher}\\
     \bottomrule
    \end{tabular}
\end{table}

\begin{table}[!htbp]
    \centering
    \small
    \caption{The key-recovery attacks on \cipherone. The full version has 12 rounds.}
    \label{tab:key-recovery}
    \setlength{\tabcolsep}{7pt}
    \renewcommand{\arraystretch}{1.1}
    \begin{tabular}{ccccc}
    \toprule
     Round & Time & Memory  & Data  &Ref.\\\midrule
    7 & $2^{133.6}$ &Negligible & $2^{124}$ CP  &Sect.~\ref{sect:key_recovery}\\
    7 & $2^{132}$  & $2^{129}$ Bytes &  $2^{124}$ CP&Sect.~\ref{sect:key_recovery}\\
    8 & $2^{137}$  & Negligible & $2^{128}$ CP & Sect.~\ref{sect:key_recovery}\\
    8 & $2^{129}$ &$2^{133}$ Bytes &  $2^{128}$ CP&Sect.~\ref{sect:key_recovery}\\
    \bottomrule
    \end{tabular}
\end{table}
\subsection{Organization of This Paper}
	This paper is organized as follows: Section~\ref{sec:pre} introduces the necessary notations, the specification of \cipherone, and a brief overview of the cryptanalytic techniques used in this work. In Section~\ref{sect:search_for_distinguisher}, we proposes a framework to explore DL distinguishers and apply it to construct distinguishers for each branch and the PRF. Section~\ref{sec:recovery} constructs new integral distinguishers based on algebraic degree analysis and designs a general framework that achieves the longest known key-recovery attacks on \cipherone to date. In addition, due to the flaws in the original linear security evaluation, Section~\ref{sect:linear_distinguisher} explores the optimization of linear-layer parameters for \textsf{Branch3}. Finally, Section~\ref{sec:conclusion} concludes the paper.

\section{Preliminaries}\label{sec:pre}
In this section, we start by providing the specification of \cipherone. Then we review the DL cryptanalysis and summarize the key concepts of integral cryptanalysis relevant to our work.

\subsection{Specification of \cipher}
\begin{figure}[htbp]
    \centering
    % ------- (a) PDF 图 -------
    \begin{subfigure}[t]{0.48\textwidth}
        \centering
        \includegraphics[width=\textwidth]{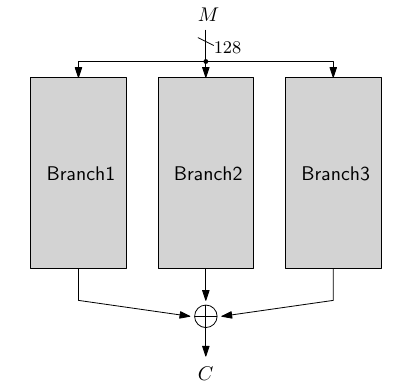}
        \label{fig:gleeok_structure}
    \end{subfigure}
    \hfill
    % ------- (b) TikZ 图 -------
    \begin{subfigure}[t]{0.48\textwidth}
        \centering
        % 等比例缩放到子图宽度
        \resizebox{\textwidth}{!}{%
        \begin{tikzpicture}[x=0.75pt,y=0.75pt,yscale=-1,xscale=1]

        %Shape: Rectangle
        \draw   (160,64.6) -- (353.6,64.6) -- (353.6,90) -- (160,90) -- cycle ;
        \draw   (160,104.6) -- (353.6,104.6) -- (353.6,130) -- (160,130) -- cycle ;
        \draw   (160,144.6) -- (353.6,144.6) -- (353.6,170) -- (160,170) -- cycle ;
        \draw   (160,184.6) -- (353.6,184.6) -- (353.6,210) -- (160,210) -- cycle ;
        \draw   (160,224.6) -- (353.6,224.6) -- (353.6,250) -- (160,250) -- cycle ;
        % 上下箭头
        \draw    (160.4,44.2) -- (160.04,62.6) ;
        \draw [shift={(160,64.6)}, rotate = 271.12] (4.37,-1.32) .. controls (2.78,-0.56) and (1.32,-0.12) .. (0,0) .. controls (1.32,0.12) and (2.78,0.56) .. (4.37,1.32)   ;
        \draw    (180,44) -- (179.64,62.4) ;
        \draw [shift={(179.6,64.4)}, rotate = 271.12] (4.37,-1.32) .. controls (2.78,-0.56) and (1.32,-0.12) .. (0,0) .. controls (1.32,0.12) and (2.78,0.56) .. (4.37,1.32)   ;
        \draw    (200,44) -- (199.64,62.4) ;
        \draw [shift={(199.6,64.4)}, rotate = 271.12] (4.37,-1.32) .. controls (2.78,-0.56) and (1.32,-0.12) .. (0,0) .. controls (1.32,0.12) and (2.78,0.56) .. (4.37,1.32)   ;
        \draw    (354,44.2) -- (353.64,62.6) ;
        \draw [shift={(353.6,64.6)}, rotate = 271.12] (4.37,-1.32) .. controls (2.78,-0.56) and (1.32,-0.12) .. (0,0) .. controls (1.32,0.12) and (2.78,0.56) .. (4.37,1.32)   ;
        \draw    (160,250) -- (159.64,268.4) ;
        \draw [shift={(159.6,270.4)}, rotate = 271.12] (4.37,-1.32) .. controls (2.78,-0.56) and (1.32,-0.12) .. (0,0) .. controls (1.32,0.12) and (2.78,0.56) .. (4.37,1.32)   ;
        \draw    (180.4,249.6) -- (180.04,268) ;
        \draw [shift={(180,270)}, rotate = 271.12] (4.37,-1.32) .. controls (2.78,-0.56) and (1.32,-0.12) .. (0,0) .. controls (1.32,0.12) and (2.78,0.56) .. (4.37,1.32)   ;
        \draw    (200.4,249.6) -- (200.04,268) ;
        \draw [shift={(200,270)}, rotate = 271.12] (4.37,-1.32) .. controls (2.78,-0.56) and (1.32,-0.12) .. (0,0) .. controls (1.32,0.12) and (2.78,0.56) .. (4.37,1.32)   ;
        \draw    (353.6,250) -- (353.24,268.4) ;
        \draw [shift={(353.2,270.4)}, rotate = 271.12] (4.37,-1.32) .. controls (2.78,-0.56) and (1.32,-0.12) .. (0,0) .. controls (1.32,0.12) and (2.78,0.56) .. (4.37,1.32)   ;
        % 虚线
        \draw  [dash pattern={on 0.84pt off 2.51pt}]  (211,54) -- (341,54) ;
        \draw  [dash pattern={on 0.84pt off 2.51pt}]  (211,260) -- (341,260) ;
        % 文本
        \draw (249,69) node [anchor=north west][inner sep=0.75pt]   [align=left] {{\fontfamily{ptm}\selectfont \textit{S}}};
        \draw (250,110.4) node [anchor=north west][inner sep=0.75pt]    {$\theta$};
        \draw (250,151.4) node [anchor=north west][inner sep=0.75pt]    {$\pi$};
        \draw (239,189.4) node [anchor=north west][inner sep=0.75pt]    {$RK_{xor}$};
        \draw (241,229.4) node [anchor=north west][inner sep=0.75pt]    {$RC_{xor}$};
        \draw (153,27) node [anchor=north west][inner sep=0.75pt]    {$x_{0}$};
        \draw (173,27) node [anchor=north west][inner sep=0.75pt]    {$x_{1}$};
        \draw (193,27) node [anchor=north west][inner sep=0.75pt]    {$x_{2}$};
        \draw (346,27) node [anchor=north west][inner sep=0.75pt]    {$x_{127}$};
        \end{tikzpicture}%
        }
        \label{fig:round_function}
    \end{subfigure}
    
    \caption{(a) Overall structure and (b) round function of \cipherone.}
    \label{fig:combined}
\end{figure}
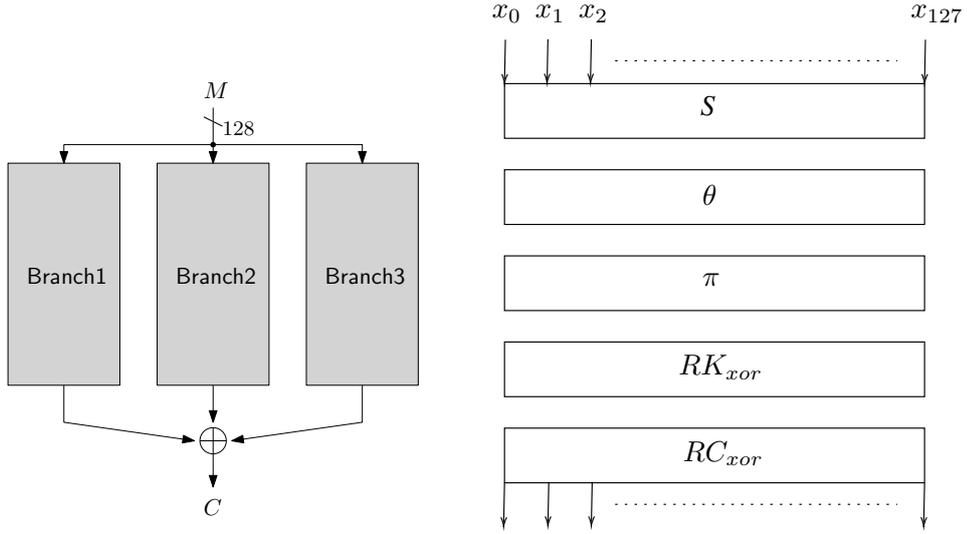

\cipher is a PRF family with two variants: \cipherone and \ciphertwo. Both variants use a 256-bit key and operate on 128-bit and 256-bit input blocks, respectively. Internally, the input message $M$ is first split and copied into three parallel branches, denoted as $\mathsf{Branch1}$, $\mathsf{Branch2}$, and $\mathsf{Branch3}$, which are processed by independent keyed permutations of 128 or 256 bits each. The final ciphertext is computed by XORing the outputs of all three branches, as shown in the following equation:
\begin{equation*}
C = \mathsf{Branch1}(X_1) \oplus \mathsf{Branch2}(X_2) \oplus \mathsf{Branch3}(X_3),
\end{equation*}
where $X_1, X_2, X_3$ are the initial internal states copied from the plaintext $M$. An overview of the overall structure is illustrated in Figure~\ref{fig:combined} (a).

Here we only introduce the detail of \cipherone and omit \ciphertwo. Each branch of \cipherone is designed as an SPN-based structure, with 12 rounds consisting of the following operations: a nonlinear layer (Sboxes), a 3-input XOR operation $\theta$, a bitwise permutation $\pi$, a round key addition $RK_{xor}$, and a round constant addition $RC_{xor}$. The round function $R$ of each branch can be represented as:
\begin{equation*}
R = RC_{xor} \circ RK_{xor} \circ \pi \circ \theta \circ S.
\end{equation*}

Figure~\ref{fig:combined} (b) shows the overview of the
round function. 
% Each operation of the round function is detailed in supplementary material. In addition, the input of each branch needs to XOR with a 128-bit whitening key before entering the first round function. 

\vspace{5pt}
\noindent\textbf{Sbox Layer ($S$).}
Different Sboxes are used in different branches. In \branchone and \branchtwo, the 3-bit Sbox $S_3$ and the 5-bit Sbox $S_5$ are applied alternately to the internal state as follows:
\[
\begin{aligned}
    X \leftarrow & (S_3(x_0||x_1||x_2)||S_5(x_3||x_4||x_5||x_6||x_7)|| \\
         & \cdots||S_3(x_{120}||x_{121}||x_{122})||S_5(x_{123}||x_{124}||x_{125}||x_{126}||x_{127})).
\end{aligned}
\]
In \branchthree, a 4-bit Sbox $S_4$ is used uniformly, and the state is updated as:
\[
\begin{aligned}
    X \leftarrow & (S_4(x_0||x_1||x_2||x_3)||S_4(x_4||x_5||x_6||x_7)|| \\
         & \cdots||S_4(x_{120}||x_{121}||x_{122}||x_{123})||S_4(x_{124}||x_{125}||x_{126}||x_{127})).
\end{aligned}
\]
The truth tables of the three Sboxes can be seen in Table~\ref{tab:sboxes} of Supplementary Material.

\vspace{5pt}
\noindent\textbf{3-input XOR ($\theta$).}
The $\theta$ operation computes each bit as a XOR of three input bits at different positions:
\[
    x_i \leftarrow x_{i+t_0}\oplus x_{i+t_1}\oplus x_{i+t_2},
\]
where the parameters of $t_{0}$, $t_{1}$, and $t_{2}$ vary across branches, summarized in Table~\ref{tab:theta-128}.

\begin{table}[!htbp]
\centering
\small
\renewcommand\tabcolsep{5pt}
\caption{The parameters of $\theta$ and $\pi$ for \cipherone}
\label{tab:theta-128}
\begin{tabular}{c|cccc}
\toprule
Param. & $t_{0}$ & $t_{1}$ & $t_{2}$ & $p$\\
\midrule
\branchone & 12 & 31 & 86 & 117\\
\branchtwo & 4  & 23 & 78 & 117 \\
\branchthree & 7  & 15 & 23 & 11\\
\bottomrule
\end{tabular}
\end{table}

\vspace{5pt}
\noindent\textbf{Permutation ($\pi$).}
The permutation $\pi$ is a branch-dependent bit-wise permutation, defined as:
\[
x_i \leftarrow x_{i \cdot p \bmod 128}.
\]
Here, $p$ denotes a branch-dependent constant, shown in Table~\ref{tab:theta-128}.

\vspace{5pt}
\noindent\textbf{Key Addition ($RK_{xor}$).}
In each round, the internal state is XORed with the round key corresponding to the current branch and round number.

\vspace{5pt}
\noindent\textbf{Constant Addition ($RC_{xor}$).}
At round $r$, the internal states of \branchone, \branchtwo, and \branchthree are XORed with the corresponding round constant $RC_{r}^i$, where $i\in\{1,2,3\}$. 
The round constants are defined as
\[
RC_{r}^i=\mathrm{MSB}_{128}\!\Big((\pi-3)\ll (r\times 128)+(i\times 12\times 128)\Big),
\]
where $\mathrm{MSB}_{128}$ returns the most significant $128$ bits. 

\vspace{5pt}
\noindent\textbf{Key Scheduling Function.}
The 256-bit master key is first divided into two 128-bit parts $K_0 = (k_0 \parallel \cdots \parallel k_{127})$ and $K_1 = (k_{128} \parallel \cdots \parallel k_{255})$.
For $\mathsf{Branch1}$, $K_0$ and $K_1$ are used directly; for $\mathsf{Branch2}$, the two halves are swapped; and for $\mathsf{Branch3}$, the halves are rotated so that $K_0 = (k_{64} \parallel \cdots \parallel k_{191})$ and $K_1 = (k_{192} \parallel \cdots \parallel k_{255} \parallel k_0 \parallel \cdots \parallel k_{63})$.
In the scheduling process $\mathrm{KSF}_{i}$ for $\mathsf{Branch}i$, as outlined in Algorithm~\ref{alg:ksf}, $K_0$ and $K_1$ are alternately applied to generate round keys. Note that the $RK_0^i$ is the whitening key of \textsf{Branch}$i$. The permutation parameter $pk_i$ is fixed to $29$, $51$, and $107$ for $\mathsf{Branch1}$, $\mathsf{Branch2}$, and $\mathsf{Branch3}$, respectively.

\begin{algorithm}
\caption{Procedure of $\mathrm{KSF}_{i}$}\label{alg-ksf}
\label{alg:ksf}
\begin{algorithmic}[1]
\State \textbf{Input:} Initial $K_0$ and $K_1$
\State \textbf{Output:} Round keys $(RK^i_{0}, RK^i_{1}, \ldots, RK^i_{R})$
\For{$r \gets 0$ \textbf{to} $R$}
    \State $(k_0 \parallel k_1 \parallel \cdots \parallel k_{127}) \gets K_{r \bmod 2}$
    \For{$j \gets 0$ \textbf{to} $127$}
        \State $k^{*}_{j} \gets k_{\,pk_{i}\cdot j \bmod 128}$
    \EndFor
    \State $K_{r \bmod 2} \gets (k^{*}_{0} \parallel k^{*}_{1} \parallel \cdots \parallel k^{*}_{127})$
    \State $RK^{i}_{r} \gets K_{r \bmod 2}$
\EndFor
\State \textbf{return} $(RK^i_{0}, RK^i_{1}, \ldots, RK^i_{R})$
\end{algorithmic}
\end{algorithm}

\subsection{Differential-Linear Cryptanalysis}

DL cryptanalysis, introduced by Langford and Hellman~\cite{DBLP:conf/crypto/LangfordH94}, combines differential and linear techniques to form a two-stage distinguisher. In the first stage, an input difference $\Delta_I$ is propagated with probability $p$ through the early rounds of the cipher. The second stage applies a linear approximation with correlation $q$ from an intermediate point to the output. When the two stages align properly in the middle of the cipher, the resulting distinguisher can be more powerful than either technique alone. Under standard assumptions, the distinguishing advantage is approximately $pq^2$, requiring $O(p^{-2}q^{-4})$ chosen plaintexts.

To improve modeling accuracy and capture dependencies between subcomponents, Bar-On et al.~\cite{DBLP:conf/eurocrypt/Bar-OnDKW19} proposed the Differential-Linear Connectivity Table (DLCT) at EUROCRYPT 2019, which is defined as
\[
    \text{DLCT}_F(\Delta, \lambda) = \#\{X\in\mathbb{F}_2^n|\lambda\cdot F(X) = \lambda\cdot F(X\oplus \Delta)\} - 2^{n-1}.
\]
\begin{figure}[htp]
\centering
\includegraphics[scale=0.6]{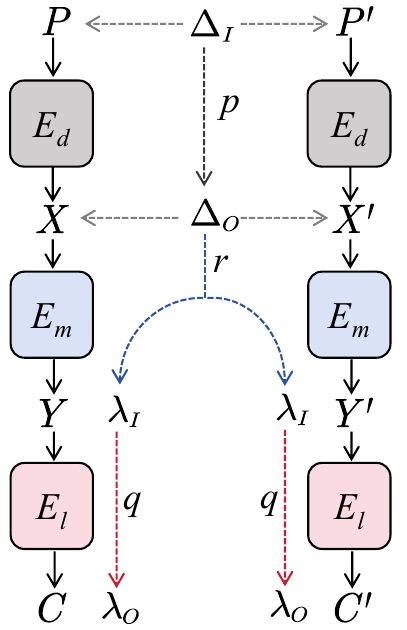}
\caption{The structure of DL distinguishers}
\label{fig:DL}
\end{figure}
Formally, the cipher is decomposed as $E = E_l \circ E_m \circ E_d$, as shown in Figure~\ref{fig:DL}, instead of the original two-stage structure. Here, $E_d$ is covered by a differential $\Delta_I \rightarrow \Delta_O$ with probability $p$ and $E_l$ is covered by a linear approximation $\lambda_I \rightarrow \lambda_O$ with correlation $q$. The correlation from $\Delta_O$ to $\lambda_I$ through the middle component $E_m$ can be computed by DLCT:
\begin{equation*}
\mathrm{C}[\Delta_O\xrightarrow{E_m}\lambda_I] = \frac{\text{DLCT}_{E_m}(\Delta_O,\lambda_I)}{2^{n-1}}= r.
\end{equation*}
Under the independence assumption between $E_d$, $E_m$, and $E_l$, the expected correlation of the DL distinguisher is evaluated by $prq^2$, and the corresponding data complexity remains $O(p^{-2}r^{-2}q^{-4})$. To be more precise, the correlation estimation, combined with the theoretical method proposed by Blondeau et al.~\cite{DBLP:journals/joc/BlondeauLN17}, can be represented as:
\begin{equation}\label{equ:DL_correlation}
    \mathrm{C}[\Delta_I\xrightarrow{E}\lambda_O] = \sum_{\Delta_O,\lambda_I}\Pr[\Delta_I\xrightarrow{E_d}\Delta_O]\cdot\mathrm{C}[\Delta_O\xrightarrow{E_m}\lambda_I]\cdot(\mathrm{C}[\lambda_I\xrightarrow{E_l}\lambda_O])^2.
\end{equation}

% In practice, the DLCT provides a more accurate estimation of the correlation by capturing the interaction between the differential and linear properties across the cipher's structure.

In practice, DL cryptanalysis often employs MILP-based methods to search for a best single DL trail, i.e.,
\begin{equation}\label{equ:DL_trail_correlation}
 \max_{\Delta_I,\Delta_O,\lambda_I,\lambda_O}\{\Pr[\Delta_I\xrightarrow{E_d}\Delta_O]\cdot\mathrm{C}[\Delta_O\xrightarrow{E_m}\lambda_I]\cdot(\mathrm{C}[\lambda_I\xrightarrow{E_l}\lambda_O])^2\},
\end{equation}
with its correlation estimated based on the DDT, LAT, and DLCT values of each involved Sbox. However, multiple trails may share a given input difference and output mask pair. According to Equation~\eqref{equ:DL_correlation}, the sum of correlations over all such trails more accurately reflects the total correlation of the distinguisher, which we refer to as the exact correlation. While this aggregation captures the combined effect of all valid trails, it still relies on specific assumptions and thus serves as a theoretical approximation. Nevertheless, for convenience, we refer to this aggregated value as the exact correlation throughout the remainder of this paper.

\subsection{Integral Cryptanalysis}

Integral cryptanalysis is a powerful technique inspired by the Square attack~\cite{DBLP:conf/fse/DaemenKR97}, and was formally introduced by Knudsen and Wagner~\cite{DBLP:conf/fse/KnudsenW02}. It exploits the behavior of a block cipher under a structured set of plaintexts to reveal statistical biases in intermediate states. Specifically, an integral distinguisher is constructed by varying a subset of input bits-called active bits-across all possible values, while fixing the remaining bits. After a few rounds, certain output bits may exhibit the \emph{balanced} property, meaning their XOR sum over all plaintexts is zero, regardless of the key.

Formally, let $E$ be an $r$-round block cipher with $n$-bit block size. Let $\mathbb{D}$ be a set of $2^d$ plaintexts differing in $d$ active bits. If the $i^{\text{th}}$ bit of the output satisfies
\[
\bigoplus_{PT \in \mathbb{D}} CT_i = 0,
\]
where $CT = E^r(PT, K)$, then bit $i$ is said to have a \emph{balanced integral property}. Such properties can be used to distinguish $r$-round $E$ from an ideal cipher.

An effective approach to constructing integral distinguishers is based on algebraic degree estimation. Suppose the algebraic degree of an $r$-round block cipher is upper bounded by $d_u$, and the number of active input bits is $d$. When $d > d_u$, the output after $r$ rounds cannot form a full-degree polynomial. In this case, the cipher has the integral distinguisher with $2^d$ chosen plaintexts. This method has been applied to evaluate ciphers such as Keccak and Luffa~\cite{DBLP:conf/fse/BouraCC11}. Several algebraic degree estimation techniques can be used for this purpose, including Boura and Canteaut's and Carlet's formulas~\cite{DBLP:journals/tit/BouraC13,DBLP:journals/tit/Carlet20a}, estimations based on the division property~\cite{DBLP:conf/eurocrypt/Todo15}, and numeric mapping~\cite{DBLP:conf/crypto/Liu17}. Among them, the division property based estimation is considered the most precise~\cite{DBLP:journals/tosc/ChenXZZ21}.

\section{Differential-Linear Cryptanalysis of \cipherone}\label{sect:search_for_distinguisher}
In~\cite{DBLP:journals/tches/AnandBCIILMRS24}, the authors give a comprehensively security analysis on \cipherone as well as all the branches from the resistance on distinguishing attacks perspective. As a consequence, the longest valid distinguisher for \cipherone is the integral distinguisher that covers 5 out of 12 rounds. Regarding the single branch, the boomerang distinguishers of \textsf{Branch1} and \textsf{Branch2} outperform the integral, linear, and differential distinguishers. The main reason is that the short differential characteristics have high probability and the probability will decrease rapidly with the round growing. Naturally, combining two short but high-probability differential characteristics can derive a good boomerang one. Inspired by this observation, we in this section will adopt another combined cryptanalytic method, DL cryptanalysis, to explore improved distinguishers for branches and the overall PRF.

\subsection{A Two-Stage Framework to Explore DL Distinguishers}
% As a derivate of the differential attack, DL cryptanalysis has shown its great power in attacking symmetric-key primitives. With the development of automatic tools for symmetric-key cryptanalysis, the automatic search for DL distinguishers and precise estimation of DL correlation also become the hot topics recently. At CRYPTO'24, Hadipour et al.~\cite{DBLP:conf/crypto/HadipourDE24} proposed an MILP/CP-aided automatic method to search for DL distinguishers by extending the DLCT layer from one round to more rounds. Meanwhile, Peng et al.~\cite{DBLP:conf/crypto/PengZWD24} presented a novel approach to accurately estimate the DL correlation based on the relation of the DL distinguisher and the truncated differential distinguisher. While both methods are highly effective, they are not directly applicable to \cipher, as its output is generated by XORing the outputs of three parallel keyed permutations, making automatic search significantly more challenging.

In this part, we introduce a two-stage framework for the automatic search of DL distinguishers. This framework is applicable to not only the single branch but also the PRF, which can be easily implemented by MILP-aided and SAT-aided tools. In the first stage, we aim to find a series of good DL trail candidates. In the second stage, for each trail, we apply a clustering-based technique to estimate its exact correlation and select the best one as our DL distinguisher. In the following context, we will illustrate in detail the framework applied to the single branch. As for its application to the PRF, we only need to combine the processes of three branches under some conditions. 

\vspace{5pt}
\noindent\textbf{Stage one: search for optimal DL trails.} In this stage, we convert the DL trail search of single branch (i.e., \branchone/\branchtwo/\branchthree in \cipherone) to an MILP\footnote{In our experiments, the MILP-based approach performed better than the SAT-based one in the second stage, e.g., computing the aggregated correlation of a 5-round DL pair of Branch1 took <1s with MILP (Gurobi) but $\approx$46s with SAT. This mainly results from Gurobi’s multi-threaded optimization and parameter tuning. Considering overall runtime and model consistency, we recommend using MILP throughout, though combining SAT for the first stage and MILP for the second may further improve efficiency at the cost of integration complexity.} problem according to Equation~\eqref{equ:DL_trail_correlation}. Specifically, we regard the single branch as an SPN-based block cipher and split the $R$-round cipher to three sub-ciphers: the top one covers $R_d$ rounds, the middle one covers only one round and the bottom sub-cipher covers $R_l$ rounds, where $R = R_d + 1 + R_l$. For the sake of simplification, we denote the three sub-ciphers by $E_d$, $E_m$ and $E_l$, respectively. Next we need to model each one separately by MILP:
\begin{itemize}
    \item The sub-model $\mathcal{M}_d$ describes the $R_d$-round differential propagation. In this sub-model, an integer variable $\mathrm{P}$ is introduced to represent the differential probability $2^{\mathrm{-P}}$; 
    \item The sub-model $\mathcal{M}_l$ describes the $R_l$-round linear mask propagation, and the involved integer variable $\mathrm{C}_l$ represents the linear approximation correlation $2^{-\mathrm{C}_l}$;
    \item The sub-model $\mathcal{M}_m$ consists of the linearly characterized DLCTs of Sbox layer and the propagation of linear mask through $\theta$ and $\pi$ operations. The integer variable $\mathrm{C}_m$ denotes the DL correlation $2^{-\mathrm{C}_m}$.
\end{itemize}
Specially, the output difference from $E_d$ propagates directly to $E_m$'s Sbox layer, the output linear mask of $\pi$ operation of $E_m$ becomes the input linear mask for $E_l$'s Sbox layer. Thus, we can integrate them to an entire model $\mathcal{M}$ by regarding $\mathcal{M}_m$ as a link of $\mathcal{M}_d$ and $\mathcal{M}_l$, describing the DL trail with round combination $(R_d,1,R_f)$. 
% As MILP modeling for differential and linear propagation in SPN block ciphers is well-established, we omit the modeling details here. The linear inequalities for characterizing the DDT, LAT, and DLCT of the 3-, 4- and 5-bit Sboxes of \cipher are provided in Supplementary Material~\ref{Supplementary Material:Ineq}.
According to Equation~\eqref{equ:DL_trail_correlation}, our aim is to maximize the product of the differential probability for $E_d$, the correlation for $E_m$, and the squared correlation for $E_l$, i.e., $2^{-\mathrm{P}}\cdot 2^{-\mathrm{C}_m}\cdot 2^{-2\mathrm{C}_l}.$
Consequently, the model's objective function is 
\[
    \text{minimize: }\mathrm{P}+\mathrm{C}_m+2\cdot\mathrm{C}_l.
\]
Send this model to an MILP solver (we in this paper use Gurobi optimization\footnote{\url{https://www.gurobi.com/}}), then an $R$-round DL trail with the best correlation under the round configuration $(R_d,1,R_l)$ can be found. 

Assuming the returned solution yields an objective function value of $\mathrm{C}$, all DL trails under the round configuration $(R_d,1,R_l)$ exhibit correlations of at most $2^{-\mathrm{C}}$.
To collect more high-quality DL trails, we constrain the objective function to $\mathrm{C}$, and iteratively add the \textit{cutting-off inequality} (see Equation~\eqref{equ:cutting_off} in Supplementary Material), which is originally proposed by Sun et al.~\cite{DBLP:conf/asiacrypt/SunHWQMS14}, to exclude previously generated difference-mask pairs. This process continues until the updated model becomes infeasible (as outlined in Algorithm~\ref{alg:enumerate_solution}), then a set of  difference-mask pairs, all with the same optimal correlation of $2^{-\mathrm{C}}$, is obtained. Next, we move to the second stage.

\begin{algorithm}[!htbp]
\caption{Collect $R$-round high-quality DL trails}\label{alg:enumerate_solution}
\small
\begin{algorithmic}[1]
\State \textbf{Input:} The round configuration $(R_d, 1, R_l)$
\State \textbf{Output:} A set of difference–mask pairs and the common correlation
\State Initialize $\mathbb{P}$ as an empty set
\State Construct an MILP model $\mathcal{M}$ as described above and solve $\mathcal{M}$ using Gurobi
\State Retrieve the correlation weight $\mathrm{C}$ and the difference–mask pair $(\Delta, \lambda)$ from $\mathcal{M}$'s current solution
\State Add $(\Delta, \lambda)$ to $\mathbb{P}$
\State Append a cutting–off inequality corresponding to $(\Delta, \lambda)$ to $\mathcal{M}$ and constrain $\mathcal{M}$'s objective function to $\mathrm{C}$
\While{$\mathcal{M}$ is feasible}
    \State Retrieve the new difference–mask pair $(\Delta, \lambda)$ from $\mathcal{M}$'s current solution
    \State Add $(\Delta, \lambda)$ to $\mathbb{P}$
    \State Append a cutting–off inequality corresponding to $(\Delta, \lambda)$ to $\mathcal{M}$
\EndWhile
\State \textbf{return} $(\mathbb{P}, 2^{-\mathrm{C}})$
\end{algorithmic}
\end{algorithm}

\vspace{5pt}
\noindent\textbf{Stage two: compute the aggregated correlations.} In this stage, we aim to compute the exact correlation for each prepared DL trail and select the one with the highest value. As specified by Equation~\eqref{equ:DL_correlation}, given a difference-mask pair $(\Delta_I, \lambda_O)$, we must enumerate all compatible $(\Delta_O, \lambda_I)$ pairs and compute their signed correlations. The exact correlation for $\Delta_I\xrightarrow{E}\lambda_O$ is then obtained by summing these signed values. Consequently, modeling the correlation sign of DL trails becomes essential. Note that the correlation sign of the trail $\Delta_I\xrightarrow{E_d}\Delta_O\xrightarrow{E_m}\lambda_I\xrightarrow{E_l}\lambda_O$ is identical to that of $\Delta_O\xrightarrow{E_m}\lambda_I$. This equivalence holds because both the differential trail $\Delta_I\xrightarrow{E_d}\Delta_O$ (contributing probability) and the linear trail $\lambda_I\xrightarrow{E_l}\lambda_O$ (contributing squared correlation) are strictly positive. Since the middle part $E_m$ covers only one round in our DL trail, the correlation sign of $\Delta_O\xrightarrow{E_m}\lambda_I$ can be directly derived from the DLCTs of the Sboxes within $E_m$. Therefore, we only require signed DLCT modeling for $E_m$. To achieve this, we re-characterize the Sbox's DLCT representation by introducing an auxiliary variable $s$ to encode the correlation sign as
\begin{equation}\label{equ:encode_correlation_sign}
    s = 
\begin{cases}
    0 & \text{if the DLCT's entry is positive,} \\
    1 & \text{if the DLCT's entry is negative.}
\end{cases}
\end{equation}
For example, Table~\ref{tab:S3DLCT} shows the DLCT of \cipher's 3-bit Sbox. All non-zero entries have an absolute value of 4, confirming that the absolute value of correlation of valid DL transition is constantly 1. Therefore, we encode the signed DLCT using 7-dimensional points. For instance, the valid DL transitions $0\textup{x}1\xrightarrow{\text{DLCT}}0\textup{x}1$ and $0\textup{x}0\xrightarrow{\text{DLCT}}0\textup{x}7$ are with entries $-4$ and 4, encoded as $(0,0,1,0,0,1,1)$ and $(0,0,0,1,1,1,0)$, respectively. Using the algorithm in \cite{Feng2022FullLI}, we obtain a system of inequalities (see Equation~\eqref{eq:DLCTS3_sgn} in Supplementary Material~\ref{Supplementary Material:Ineq}) characterizing the signed DLCT through the MILP-based inequality selection method. The sign (positive or negative) of a trail's correlation is determined by the parity (even or odd) of the number of Sboxes that have a sign of 1. An even count yields positive correlation, and an odd count yields negative correlation. Therefore, if we use a binary variable $\mathrm{s}$ to represent the sign of a trail's correlation as
\[
    \mathrm{s} = \begin{cases}
    0 & \text{if the trail's correlation is positive,} \\
    1 & \text{otherwise,}
\end{cases}
\]
then $\mathrm{s}$ can be constrained by
\[
    \mathrm{s} + \sum_{i=1}^{m} s_i = 2d, 
\]
where $m$ ($m=32$ for \cipherone) is the number of Sboxes, $s_i\in\{0,1\}$ is the sign of the $i^{\text{th}}$ Sbox's correlation and $d\in \mathbb{Z}$ is a dummy integer variable ranging from $0$ to $m/2$.

\begin{figure}[htp]
\centering
\includegraphics[scale=0.48]{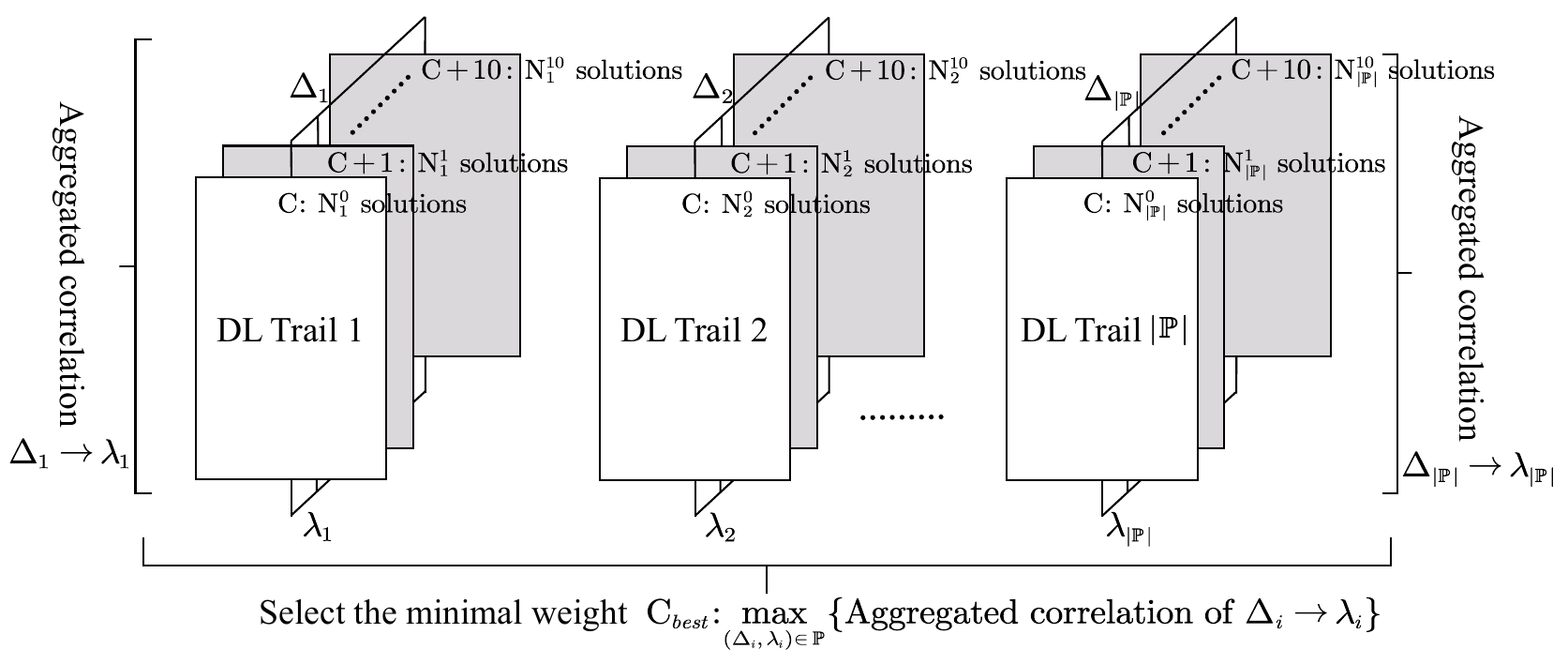}
\caption{Overview of Algorithm~\ref{alg:second_stage}}
\label{fig:second_stage}
\end{figure}

Based on the signed correlation modeling method, we now proceed to compute the exact correlation for a given difference-mask pair. Our core idea is to enumerate all valid trails sharing this common pair under a fixed correlation weight and compute the aggregation effect by summing their signed correlations. We assume that the first stage search (Algorithm~\ref{alg:enumerate_solution}) has already provided a minimum correlation weight $\mathrm{C}$ and a set of difference-mask pairs $\mathbb{P}$. For a pair $(\Delta, \lambda) \in \mathbb{P}$, we aim to compute its aggregated correlation by reusing the MILP model $\mathcal{M}$ constructed in the first stage for finding the best DL trail. To adapt $\mathcal{M}$ for this purpose, we replace the unsigned DLCT constraints in the middle part $E_m$ with their signed counterparts, as described earlier. We then fix the input difference and output mask to $\Delta$ and $\lambda$, respectively, and constrain the model's objective function to a correlation weight $\mathrm{C}' \geq \mathrm{C}$. Finally, we configure the solver parameters\footnote{For more details, please refer to Gurobi's reference manual at \url{https://docs.gurobi.com/projects/optimizer/en/current/}.} to enumerate all feasible trails under these constraints as below:
\[
    \mathcal{M}.\textsf{Params.PoolSearchMode} = 2
\]
and 
\[
    \mathcal{M}.\textsf{Params.PoolSolutions} = 200000000.
\]
This parameter configuration enables Gurobi to generate a solution pool when solving the model $\mathcal{M}$. Each solution corresponds to a unique trail with difference-mask pair $(\Delta, \lambda)$ and signed correlation $(-1)^{\mathrm{s}}\cdot2^{-\mathrm{C}'}$. Given $\mathrm{N}$ total solutions in the pool, we compute the aggregated signed correlation for weight $\mathrm{C}'$ as
\[
    2^{-\mathrm{C'}}\cdot\sum_{i=1}^{\mathrm{N}} (-1)^{\mathrm{s}_i},
\]
where $\mathrm{s}_i$ denotes the correlation sign of the trail corresponding to the $i^{\text{th}}$ solution.
By repeating this process for correlation weights from $\mathrm{C}$ down to a threshold weight $\mathrm{C}_t$ (typically with $\mathrm{C}_t - \mathrm{C}\geq 10$), we approximate the exact correlation of the DL distinguisher $\Delta\xrightarrow{E}\lambda$ through the summation:
\[
    \sum_{\mathrm{C}' = \mathrm{C}}^{\mathrm{C}_t} \left(2^{-\mathrm{C}'}\cdot\sum_{i}(-1)^{\mathrm{s}_i}\right),
\]
where the inner sum is taken over all trails with correlation weight $\mathrm{C}'$. By computing and comparing the aggregated correlation for each pair in $\mathbb{P}$, we select the distinguisher with the maximum absolute correlation as the final DL distinguisher. The overall procedure of the second stage is outlined in Algorithm~\ref{alg:second_stage} and a schematic illustration is provided in Figure~\ref{fig:second_stage}.
\begin{algorithm}[!h]
\caption{Select the best one from the prepared DL trails}\label{alg:second_stage}
\begin{algorithmic}[1]
\State \textbf{Input:} The minimal correlation weight $\mathrm{C}$, a set of difference-mask pairs $\mathbb{P}$, and an MILP model $\mathcal{M}$ that describes $(R_d,1,R_l)$-round DL trail.
\State \textbf{Output:} A difference-mask pair and its aggregated correlation.

\State $\mathrm{C}_{best} \gets 0$, $\mathrm{C}_t \gets \mathrm{C} + 10$, $(\Delta_I, \lambda_O) \gets (0,0)$
\For{each $(\Delta, \lambda)$ in $\mathbb{P}$}
    \State $\mathrm{C}_{temp} \gets 0$
    \State Re-constrain $\mathcal{M}$'s difference-mask pair to $(\Delta, \lambda)$
    \For{$\mathrm{C'}$ from $\mathrm{C}$ to $\mathrm{C}_t$}
        \State Re-constrain $\mathcal{M}$'s objective function to $\mathrm{C'}$
        \State $\mathcal{M}.\textsf{Params.PoolSearchMode} \gets 2$
        \State $\mathcal{M}.\textsf{Params.PoolSolutions} \gets 200000000$
        \State Solve $\mathcal{M}$ by Gurobi
        \State $\mathrm{N} \gets \mathcal{M}.\textsf{SolCount}$
        \State $\mathrm{s}_{temp} \gets 0$
        \For{$i$ from $0$ to $\mathrm{N}-1$}
            \State $\mathcal{M}.\textsf{Params.SolutionNumber} \gets i$
            \State Retrieve the correlation sign $\mathrm{s}_i$ from the $i^{\text{th}}$ solution
            \State $\mathrm{s}_{temp} \gets \mathrm{s}_{temp} + (-1)^{\mathrm{s}_i}$
        \EndFor
        \State $\mathrm{C}_{temp} \gets \mathrm{C}_{temp} + \mathrm{s}_{temp} \cdot 2^{-\mathrm{C'}}$
    \EndFor
    \If{$|\mathrm{C}_{temp}| > \mathrm{C}_{best}$}
        \State $\mathrm{C}_{best} \gets |\mathrm{C}_{temp}|$
        \State $(\Delta_I, \lambda_O) \gets (\Delta, \lambda)$
    \EndIf
\EndFor
\State \Return $(\Delta_I, \lambda_O)$ and $\mathrm{C}_{best}$
\end{algorithmic}
\end{algorithm}

\vspace{5pt}
\noindent\textbf{Application to the overall PRF.} As previously stated, the above two-stage framework can be applied not only to branches (\branchone/\branchtwo/\branchthree) but also to the overall PRF. For the PRF setting, in the first stage, we establish a common round configuration and construct models for each of the three inner branches individually. These models are then integrated into a unified model by constraining them to share identical input difference and output mask, according to the differential propagation rule through the branching operation as well as linear mask propagation rule through the XOR operation. Furthermore, the objective function of this model is to minimize the sum of correlation weights of the three branches. Solving this model with Gurobi provides a set of difference-mask pairs, the minimum correlation weight for PRF as well as the correlation weights for each of the three inner branches, as outlined in Algorithm~\ref{alg:enumerate_solution}. 
    In the second stage, we compute the aggregated correlations for each branch individually (Lines 5-18, Algorithm~\ref{alg:second_stage}), and subsequently derive the overall aggregated correlation for the PRF using the Pilling-up Lemma~\cite{DBLP:conf/eurocrypt/Matsui93}.

\subsection{Searching for DL Distinguishers of \cipherone}\label{sect:DL-distinguisher}
We apply the two-stage framework to search for DL distinguishers of \cipherone and its underlying keyed permutations \branchone/\branchtwo/\branchthree. All the DDTs, LATs, DLCTs as well as the corresponding linear inequality descriptions used in the search can be found in Supplementary Material.

\vspace{5pt}
\noindent\textbf{Setting round configuration.} The crucial step before searching is to choose an appropriate round configuration. Inspired by the differential and linear behaviors of each branches (see Table~\ref{tab:diff_linear_branch1_2}), we give the following round configuration: 
\[
(R_d,1,R_l) = 
\begin{cases}
    (\frac{R-1}{2},1,\frac{R-1}{2}) & \text{if $R$ is odd,}\\
    (\frac{R}{2}-1,1,\frac{R}{2}) & \text{otherwise,}
\end{cases}
\]
for \branchone/\branchtwo/\branchthree since both the differential probability and linear approximation squared correlation decrease rapidly with the rounds growing, while the latter declines at a slightly slower rate compared to the former. For instance, to search for 5-round DL distinguisher of \branchone, we prefer the configuration $(2,1,2)$ rather than $(3,1,1)$ or $(1,1,3)$ since the former one can probably derive the DL trail with a correlation of $2^{-16}$ whereas that of the latter exhibits $2^{-26}$ or $2^{-24}$.

\begin{table}[!htbp]
\centering
\small
\caption{The optimal differential and linear distinguishers~\cite{DBLP:journals/tches/AnandBCIILMRS24} up to five rounds for \branchone, \branchtwo, and \branchthree. DC and C$^2$ denote the differential probability and squared correlation, respectively.\label{tab:diff_linear_branch1_2}}
\renewcommand{\tabcolsep}{6pt}
\begin{tabular}{@{}c|c|ccccc@{}}
\toprule
\multirow{2}{*}{Branch} & \multirow{2}{*}{DC/C$^2$} & \multicolumn{5}{c}{\#Round} \\ 
\cmidrule(lr){3-7}
 & & 1 & 2 & 3 & 4 & 5 \\ 
\midrule
\multirow{2}{*}{\branchone} & DC & $2^{-2}$ & $2^{-8}$ & $2^{-24}$ & $2^{-50}$ & $\leq 2^{-67}$ \\
& C$^2$ & $2^{-2}$ & $2^{-8}$ & $2^{-22}$ & $2^{-44}$ & $\leq 2^{-64}$ \\ 
\midrule
\multirow{2}{*}{\branchtwo} & DC & $2^{-2}$ & $2^{-8}$ & $2^{-24}$ & $2^{-50}$ & $\leq 2^{-66}$ \\
& C$^2$ & $2^{-2}$ & $2^{-8}$ & $2^{-22}$ & $2^{-44}$ & $\leq 2^{-64}$ \\ 
\midrule
\multirow{2}{*}{\branchthree} & DC & $2^{-2}$ & $2^{-8}$ & $2^{-20}$ & $2^{-32}$ & $2^{-36}$ \\
& C$^2$ & $2^{-2}$ & $2^{-8}$ & $2^{-20}$ & $2^{-30}$ & $2^{-36}$ \\ 
\bottomrule
\end{tabular}
\end{table}

\interfootnotelinepenalty=10000

\vspace{5pt}
\noindent\textbf{Finding DL distinguishers.} Based on the aforementioned round configurations and utilizing our two-stage algorithms, we identify effective DL distinguishers for \cipherone and its underlying branches. In particular, we explore DL distinguishers up to 7/7/8 rounds for its \sloppy\branchone/\branchtwo/\branchthree with the squared correlation\footnote{It means that the square of the aggregated correlation of DL distinguishers, which can be intuitively compared to other distinguishing attacks in terms of the data complexity. For instance, the 7-round DL distinguisher of \branchone with squared correlation $2^{-88.12}$ indicates that distinguishing \branchone from a random permutation requires approximately $2^{88.12}$ pairs of chosen plaintexts.} $2^{-88.12}$/$2^{-88.12}$/$2^{-38.73}$. Note that the best known distinguisher for \branchone/\branchtwo/\branchthree, presented by designers~\cite{DBLP:journals/tches/AnandBCIILMRS24}, is the 7/7/8-round boomerang/boomerang/differential distinguisher with probability $2^{-100}$/$2^{-100}$/$2^{-64}$. Compared to these results, our DL distinguishers do not cover a larger number of rounds, but the data complexities are significantly reduced. As for the whole PRF \cipherone, we can only find DL distinguishers up to 4 rounds limited to the computation resource. We list the result in Table~\ref{tab:branch_dl}.
\begin{table}[!htbp]
    \centering
    \small
    \caption{The squared correlation of DL distinguishers for \cipherone.}
    \label{tab:branch_dl}
    \setlength{\tabcolsep}{4pt}
    \begin{tabular}{c|cccccccc}
    \toprule
     \multirow{2}{*}{Cipher} & \multicolumn{8}{c}{\#Round}\\\cmidrule{2-9}
         &  1 & 2 & 3 & 4 & 5 & 6 & 7 & 8\\\midrule
     \textsf{Branch1} & 1 & 1 & 1 & $2^{-5.36}$ & $2^{-18.86}$ & $2^{-44.83}$ & $2^{-88.12}$ & -\\
     \textsf{Branch2} & 1 & 1 & 1 & $2^{-5.36}$ & $2^{-18.86}$ & $2^{-44.83}$ & $2^{-88.12}$ & -\\
     \textsf{Branch3} & 1 & 1 & $2^{-0.83}$ & $2^{-5.77}$ & $2^{-12.42}$ & $2^{-22.00}$ & $2^{-28.90}$ & $2^{-38.73}$\\
     \cipherone & 1 &$2^{-0.39}$ & $2^{-11.72}$ & $2^{-49.04}$ &$\leq 2^{-127.40}$ & - & - & -\\
     \bottomrule
    \end{tabular}
\end{table}

It is observed from our experimental results that there is only one active Sbox in the input difference as well as the output mask of the high-quality DL trails in general. To investigate the upper bound of 5-round distinguisher, we separately compute the aggregated correlation for three branches by constraining the number of Sbox of input/output to one. The results are $2^{-23}$, $2^{-23}$ and $2^{-17.70}$, respectively. Note that we do not constrain that the input/output of the three models are identical to each other. Consequently, we regard the product of the three aggregated correlations as the upper bound, i.e., $2^{-63.70}$.

\vspace{5pt}
\noindent\textbf{Verification on DL distinguishers.} All the above DL distinguishers are explored and evaluated by leveraging MILP automatic tool based on our two-stage framework. To validate the correctness of these DL distinguishers and prove the effectiveness of our framework, we conduct some experimental verification on branches as well as the PRF under 100 random keys and record the average value. The results are listed in Table~\ref{tab:experiments}. It can be seen that the gap between the estimated values and experimental values is notably small, primarily attributable to statistical bias, which is almost negligible. These results demonstrate both the validity of the high-round DL distinguishers we identified through our search and the efficacy of our methodology. 

\begin{table}[!htbp]
    \centering
    % \small
    \caption{Verification on DL distinguishers of \textsf{\cipher-128} and its underlying branches. Est. and Exp. denote the estimated value and experimental value, respectively.}
    \label{tab:experiments}
    \setlength{\tabcolsep}{3.5pt}
    \begin{tabular}{c|c|c|c|c}
    \toprule
     Cipher & Round & Difference-mask pair & Est. & Exp. \\\midrule
     \branchone & 5 & \makecell{(0x6000,\\ 0x89df5aa33e89239079ebd0c7d964685d)} & $2^{-18.86}$ & $2^{-18.85}$ \\\midrule
     \branchtwo & 5 & \makecell{(0xe00000000000000000000000,\\0x7e656d6aacfa669e41e7a7431f659180)} & $2^{-18.86}$ & $2^{-18.87}$ \\\midrule
     \branchthree & 6 & \makecell{(0x800000000000000,\\0x40000040000000000000000)} & $2^{-22.00}$ & $2^{-21.52}$ \\\midrule
     \cipherone  & 3 &\makecell{(0x200000000000000000000,\\0x200000000000000000000000000)} & $2^{-11.72}$ & $2^{-11.70}$  \\
     \bottomrule
    \end{tabular}
\end{table}

\section{Integral-Based Key-Recovery Attacks on \cipherone}\label{sec:recovery}
In this section, we present the key-recovery attacks on \cipherone. We start by identifying new integral distinguishers for its three branches, based on an analysis of the upper bound on algebraic degree. Notably, the number of rounds covered by the integral distinguishers differs between \branchone/\branchtwo and \branchthree. Using this observation, we then design a general attack framework by prepending one round before the distinguisher to enable key recovery. Following this, we apply this framework in two settings: with the full codebook and without the full codebook.

\subsection{New Integral Distinguishers for \cipherone}\label{sect:integral_distinguisher}
In~\cite{DBLP:journals/tches/AnandBCIILMRS24}, the authors employ the SAT-based automatic tool based on division property to search for integral distinguishers of \cipherone's underlying branches. Under the data complexity of $2^{127}$, they obtain the distinguishers up to 6/6/5/5 rounds for \branchone/\branchtwo/\branchthree/\cipherone. However, for more rounds, determining the existence of such distinguishers becomes computationally infeasible within a reasonable time. To address this limitation and enable efficient exploration of longer-round distinguishers, we instead adopt an approach based on evaluating the upper bound of the algebraic degree. 

To demonstrate the effectiveness of degree evaluation in identifying integral distinguishers, let us first revisit the data complexity of the 6-/6-round integral distinguisher~\cite{DBLP:journals/tches/AnandBCIILMRS24} of \branchone/\branchtwo. Note the nonlinear layer of \branchone/\branchtwo adopts the 3- and 5-bit variants of the $\chi$ operations, resulting an algebraic degree of 2 for its round function. Given this, the trivial upper bound on the algebraic degree of 6-round output is $2^6 = 64$. Therefore, by activating 65 input bits, one can construct a structure of $2^{65}$ plaintexts that guarantees all output bits are balanced after 6 rounds. This leads to a valid 6-round integral distinguisher. Compared to the integral distinguisher proposed in the original design document~\cite{DBLP:journals/tches/AnandBCIILMRS24}, our newly explored one based on degree evaluation significantly reduced the data complexity (from $2^{127}$ to $2^{65}$).

In order to construct longer integral distinguishers, it is essential to tighten the upper bound on the algebraic degree. To this end, we adopt the iterative degree estimation formula proposed by Carlet, as illustrated in the following theorem, to estimate the algebraic degree of \branchone/\branchtwo/\branchthree for rounds exceeding 6/6/4, respectively.
\begin{theorem}[Carlet-bound~\cite{DBLP:journals/tit/Carlet20a}]
Let $F$ be a permutation of $\mathbb{F}_2^n$ and let $G$ be a function from $\mathbb{F}_2^n$ to $\mathbb{F}_2^m$. Then we have
\[
    \deg(G\circ F)\leq n - \left \lceil \frac{n-\deg(G)}{\deg(F^{-1})} \right \rceil,
\]
where $F^{-1}$ represents the inverse of $F$.
\end{theorem}
We give the following example to clarify how to use the Carlet-bound to evaluate the algebraic degree.

\begin{example}
    We have previously estimated the algebraic degree for 6-round \branchone/\branchtwo according to the trivial bound. To evaluate the 7-round case, we treat the first 6 rounds as a function $G$ and the $7^{\text{th}}$ round encryption as the function $F$, thus $\deg(G) = 64$ and $\deg(F^{-1}) = 3$ as the inverses of 3-bit Sbox and 5-bit Sbox have the algebraic degrees of 2 and 3, respectively. According to the theorem, the 7-round algebraic degree, $\deg(G\circ F)$, can be derived: 
\[
    128 - \left \lceil \frac{128-64}{3} \right \rceil = 106.
\]
\end{example}

By iteratively applying Carlet-bound, we obtain the upper bound on algebraic degree for \branchone/\branchtwo/\branchthree. Moreover, the degree of the overall PRF \cipherone can be determined by the maximum degree among the underlying branches. The results are summarized in Table~\ref{tab:degree}. As shown, these bounds confirm the existence of longer-round integral distinguishers. To be more specific, the 9-/9-/7-/7-round integral distinguisher can be constructed for \branchone/\branchtwo/\branchthree/\cipherone with a data complexity $2^{126}$/$2^{126}$/$2^{127}$/$2^{127}$, such that all the output bits are balanced.
\begin{table}[!htbp]
    \centering
    \small
    \caption{The upper bound on algebraic degree of \branchone/\branchtwo/\branchthree as well as \cipherone up to 10 rounds.}
    \label{tab:degree}
    \setlength{\tabcolsep}{5pt}
    \begin{tabular}{c|cccccccccc}
    \toprule
     \multirow{2}{*}{Cipher} & \multicolumn{10}{c}{\#Round}\\\cmidrule{2-11}
         &  1 & 2 & 3 & 4 & 5 & 6 & 7 & 8 & 9 & 10\\\midrule
     \branchone & 2 & 4 & 8 & 16 & 32 & 64 & 106 & 120 & 125 & 127\\
     \branchtwo & 2 & 4 & 8 & 16 & 32 & 64 & 106 & 120 & 125 & 127\\
     \branchthree & 3 & 9 & 27 & 81 & 112 & 122 & 126 & 127 & 127 & 127 \\
     \cipherone & 3 & 9 & 27 & 81 & 112 & 122 & 126 & 127 & 127 & 127 \\
     \bottomrule
    \end{tabular}
\end{table}

\subsection{An Integral-Based Attack Framework}
In the general key-recovery attacks based on integral distinguishers, it needs to prepend or/and append a certain number of rounds to make guesses about
the related key bits that satisfy the distinguisher. However, for \cipherone, which consists of three branches, the adversary need to guess
two outputs of the permutations to perform the backward computation from the output. This structural feature significantly complicates the attachment of key-recovery steps at the end of the cipher.
For instance, consider appending one additional round (excluding the linear layer) to a 4-round integral distinguisher of \branchone, which can be mounted with a data complexity of $2^{17}$. To detect the integral property of one output bit of 4-round \branchone, the attacker must guess 4 related key bits and 4 bits of sum of 5-round outputs of the two other branches. Consequently, for each key guess, the adversary would need to evaluate $(2^4)^{2^{17}} = 2^{2^{19}}$ potential values---far exceeding the $2^{256}$ complexity of an exhaustive key search. 

To avoid guessing the outputs of the other two branches, we can only prepend rounds before the distinguisher. 
It is observed from Table~\ref{tab:degree} that \branchone and \branchtwo exhibit a lower degree growth compared to \branchthree, which implies that prepending one or two rounds to \branchone/\branchtwo does not affect their output balance property under the same data complexity. Therefore, we only need to guess the key bits involved in the prepended rounds of \branchthree.
For example, the 4-round distinguisher of \cipherone requires a data complexity of $2^{82}$. After prepending one/two rounds before each branches, the output of 5-/6-round \branchone/\branchtwo still remains balanced under data complexity $2^{82}$ as the degree of 5-/6-round \branchone/\branchtwo is no more than 32/64. Thus we only need to guess the related-key bits in the prepended rounds of \branchthree.

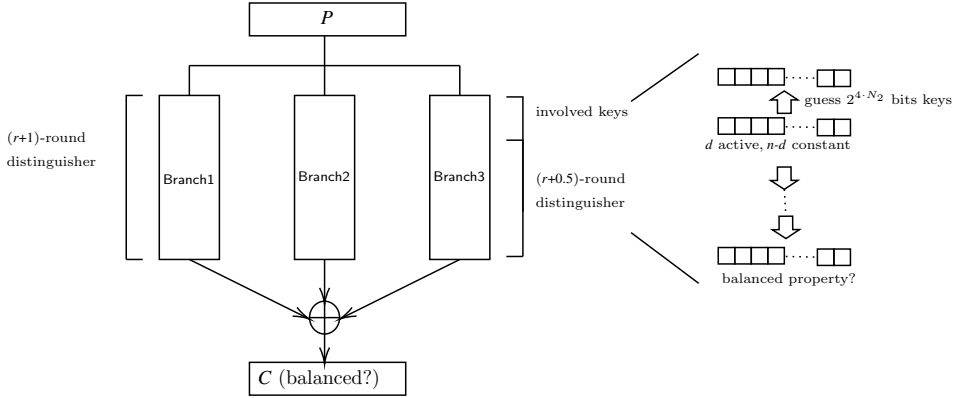
\begin{figure}[htp]
\tikzset{every picture/.style={line width=0.75pt}} %set default line width to 0.75pt        
 \vspace{1em}  % 根据图形高度微调，给正文留白

\centering
\tikzset{every picture/.style={line width=0.75pt}} %set default line width to 0.75pt        
\begin{center}
\begin{adjustbox}{center,scale=0.75} 
\

\tikzset{every picture/.style={line width=0.75pt}} %set default line width to 0.75pt        

\begin{tikzpicture}[x=0.75pt,y=0.75pt,yscale=-1,xscale=1]

%Shape: Rectangle [id:dp06949997767605365] 
\draw   (164,11) -- (268,11) -- (268,33) -- (164,33) -- cycle ;
%Straight Lines [id:da9630988207117113] 
\draw    (214,33) -- (214,53) ;
%Straight Lines [id:da12309055173743633] 
\draw    (124,53) -- (304,53) ;
%Straight Lines [id:da5657934238166553] 
\draw    (124,53) -- (124,73) ;
%Shape: Rectangle [id:dp9238364236428491] 
\draw   (104,73) -- (144,73) -- (144,183) -- (104,183) -- cycle ;
%Shape: Rectangle [id:dp20642755494513287] 
\draw   (194,73) -- (234,73) -- (234,183) -- (194,183) -- cycle ;
%Straight Lines [id:da17506360661131004] 
\draw    (214,53) -- (214,73) ;
%Shape: Rectangle [id:dp3778933727561822] 
\draw   (284,73) -- (324,73) -- (324,183) -- (284,183) -- cycle ;
%Straight Lines [id:da7973421025342082] 
\draw    (304,53) -- (304,73) ;
\draw   (204,222) .. controls (204,215.92) and (208.48,211) .. (214,211) .. controls (219.52,211) and (224,215.92) .. (224,222) .. controls (224,228.08) and (219.52,233) .. (214,233) .. controls (208.48,233) and (204,228.08) .. (204,222) -- cycle ; \draw   (204,222) -- (224,222) ; \draw   (214,211) -- (214,233) ;
%Straight Lines [id:da35009221689964076] 
\draw    (124,183) -- (202.21,222.11) ;
\draw [shift={(204,223)}, rotate = 206.57] [color={rgb, 255:red, 0; green, 0; blue, 0 }  ][line width=0.75]    (10.93,-3.29) .. controls (6.95,-1.4) and (3.31,-0.3) .. (0,0) .. controls (3.31,0.3) and (6.95,1.4) .. (10.93,3.29)   ;
%Straight Lines [id:da05397958629329591] 
\draw    (304,183) -- (225.79,222.11) ;
\draw [shift={(224,223)}, rotate = 333.43] [color={rgb, 255:red, 0; green, 0; blue, 0 }  ][line width=0.75]    (10.93,-3.29) .. controls (6.95,-1.4) and (3.31,-0.3) .. (0,0) .. controls (3.31,0.3) and (6.95,1.4) .. (10.93,3.29)   ;
%Straight Lines [id:da18836875199831837] 
\draw    (214,183) -- (214,211) ;
\draw [shift={(214,213)}, rotate = 270] [color={rgb, 255:red, 0; green, 0; blue, 0 }  ][line width=0.75]    (10.93,-3.29) .. controls (6.95,-1.4) and (3.31,-0.3) .. (0,0) .. controls (3.31,0.3) and (6.95,1.4) .. (10.93,3.29)   ;
%Straight Lines [id:da5035786968991518] 
\draw    (214,233) -- (214,251) ;
\draw [shift={(214,253)}, rotate = 270] [color={rgb, 255:red, 0; green, 0; blue, 0 }  ][line width=0.75]    (10.93,-3.29) .. controls (6.95,-1.4) and (3.31,-0.3) .. (0,0) .. controls (3.31,0.3) and (6.95,1.4) .. (10.93,3.29)   ;
%Shape: Rectangle [id:dp828847442751026] 
\draw   (164,252) -- (268,252) -- (268,274) -- (164,274) -- cycle ;
%Shape: Right Angle [id:dp5038642159186411] 
\draw   (335,103) -- (346,103) -- (346,181) ;
%Straight Lines [id:da8414386477417348] 
\draw    (335,181) -- (346,181) ;
%Shape: Right Angle [id:dp8779859230797538] 
\draw   (335,74) -- (346,74) -- (346,151) ;
%Shape: Right Angle [id:dp8651176863791905] 
\draw   (93,73) -- (82,73) -- (82,183) ;
%Straight Lines [id:da32757673335522] 
\draw    (82,183) -- (93,183) ;
%Straight Lines [id:da45976607674896364] 
\draw    (418,77) -- (463,45) ;
%Straight Lines [id:da4276047468760944] 
\draw    (418,165) -- (463,199) ;
%Shape: Rectangle [id:dp3577773513864674] 
\draw   (476,88) -- (487,88) -- (487,99) -- (476,99) -- cycle ;
%Shape: Rectangle [id:dp13158105158365763] 
\draw   (487,88) -- (498,88) -- (498,99) -- (487,99) -- cycle ;
%Shape: Rectangle [id:dp8639567535820212] 
\draw   (498,88) -- (509,88) -- (509,99) -- (498,99) -- cycle ;
%Shape: Rectangle [id:dp2885675922190615] 
\draw   (509,88) -- (520,88) -- (520,99) -- (509,99) -- cycle ;
%Straight Lines [id:da4979713301080936] 
\draw  [dash pattern={on 0.84pt off 2.51pt}]  (521,94) -- (540.6,94.4) ;
%Shape: Rectangle [id:dp35170445592408306] 
\draw   (542,89) -- (553,89) -- (553,100) -- (542,100) -- cycle ;
%Shape: Rectangle [id:dp6262433369445535] 
\draw   (553,89) -- (564,89) -- (564,100) -- (553,100) -- cycle ;
%Shape: Rectangle [id:dp4605476411768802] 
\draw   (476,175) -- (487,175) -- (487,186) -- (476,186) -- cycle ;
%Shape: Rectangle [id:dp18208261453018149] 
\draw   (487,175) -- (498,175) -- (498,186) -- (487,186) -- cycle ;
%Shape: Rectangle [id:dp6272143106538433] 
\draw   (498,175) -- (509,175) -- (509,186) -- (498,186) -- cycle ;
%Shape: Rectangle [id:dp27378461142338084] 
\draw   (509,175) -- (520,175) -- (520,186) -- (509,186) -- cycle ;
%Straight Lines [id:da0638951020605194] 
\draw  [dash pattern={on 0.84pt off 2.51pt}]  (521,181) -- (540.6,181.4) ;
%Shape: Rectangle [id:dp6810826026313966] 
\draw   (542,176) -- (553,176) -- (553,187) -- (542,187) -- cycle ;
%Shape: Rectangle [id:dp015756097638190525] 
\draw   (553,176) -- (564,176) -- (564,187) -- (553,187) -- cycle ;
%Right Arrow [id:dp45917801089838184] 
\draw   (524.44,120.98) -- (524.49,129.98) -- (529.02,129.95) -- (520,136) -- (510.91,130.06) -- (515.43,130.03) -- (515.38,121.04) -- cycle ;
%Right Arrow [id:dp46321475676808144] 
\draw   (525.53,153.97) -- (525.58,162.97) -- (530.11,162.94) -- (521.09,168.99) -- (511.99,163.05) -- (516.52,163.02) -- (516.47,154.03) -- cycle ;
%Straight Lines [id:da44531121870569723] 
\draw  [dash pattern={on 0.84pt off 2.51pt}]  (520,136) -- (521,154) ;
%Shape: Rectangle [id:dp11870060225532919] 
\draw   (476,55) -- (487,55) -- (487,66) -- (476,66) -- cycle ;
%Shape: Rectangle [id:dp13653612244260016] 
\draw   (487,55) -- (498,55) -- (498,66) -- (487,66) -- cycle ;
%Shape: Rectangle [id:dp3113834201975366] 
\draw   (498,55) -- (509,55) -- (509,66) -- (498,66) -- cycle ;
%Shape: Rectangle [id:dp41457391129262333] 
\draw   (509,55) -- (520,55) -- (520,66) -- (509,66) -- cycle ;
%Straight Lines [id:da8814162929542966] 
\draw  [dash pattern={on 0.84pt off 2.51pt}]  (521,61) -- (540.6,61.4) ;
%Shape: Rectangle [id:dp2206610087169103] 
\draw   (542,56) -- (553,56) -- (553,67) -- (542,67) -- cycle ;
%Shape: Rectangle [id:dp999419349247292] 
\draw   (553,56) -- (564,56) -- (564,67) -- (553,67) -- cycle ;
%Right Arrow [id:dp37220934071384404] 
\draw   (515.44,85.01) -- (515.42,76.02) -- (510.89,76.02) -- (519.94,70.01) -- (529.01,75.99) -- (524.48,76) -- (524.5,84.99) -- cycle ;

% Text Node
\draw (210,15) node [anchor=north west][inner sep=0.75pt]   [align=left] {\textit{{\fontfamily{ptm}\selectfont P}}};
% Text Node
\draw (105,115) node [anchor=north west][inner sep=0.75pt]   [align=left] {\begin{minipage}[lt]{25.34pt}\setlength\topsep{0pt}
\begin{center}
{\fontsize{7pt}{8pt}\selectfont  \branchone}
\end{center}

\end{minipage}};
% Text Node
\draw (195,113) node [anchor=north west][inner sep=0.75pt]   [align=left] {\begin{minipage}[lt]{25.34pt}\setlength\topsep{0pt}
\begin{center}
{\fontsize{7pt}{8pt}\selectfont  \branchtwo}
\end{center}

\end{minipage}};
% Text Node
\draw (285,113) node [anchor=north west][inner sep=0.75pt]   [align=left] {\begin{minipage}[lt]{25.34pt}\setlength\topsep{0pt}
\begin{center}
{\fontsize{7pt}{8pt}\selectfont  \branchthree}
\end{center}

\end{minipage}};
% Text Node
\draw (168,255) node [anchor=north west][inner sep=0.75pt]   [align=left] {{\fontfamily{ptm}\selectfont \textit{C}} (balanced?)};
% Text Node
\draw (353,123) node [anchor=north west][inner sep=0.75pt]   [align=left] {{\footnotesize ({\fontfamily{ptm}\selectfont \textit{r}+0.5})-round }\\{\footnotesize distinguisher}};
% Text Node
\draw (353,79) node [anchor=north west][inner sep=0.75pt]   [align=left] {{\footnotesize involved keys}};
% Text Node
\draw (1,96) node [anchor=north west][inner sep=0.75pt]   [align=left] {{\footnotesize ({\fontfamily{ptm}\selectfont \textit{r}+1})-round }\\{\footnotesize distinguisher}};
% Text Node
\draw (477,191) node [anchor=north west][inner sep=0.75pt]   [align=left] {{\footnotesize balanced property?}};
% Text Node
\draw (466,101) node [anchor=north west][inner sep=0.75pt]   [align=left] {{\footnotesize {\fontfamily{ptm}\selectfont \textit{d}} active,{\fontfamily{ptm}\selectfont \textit{ n-d}} constant}};
% Text Node
\draw (532,68.4) node [anchor=north west][inner sep=0.75pt]  [font=\footnotesize]  {${\textstyle \text{guess} \ 2^{4\cdot N_{2}} \ \text{bits\ keys}}$};

\end{tikzpicture}

\end{adjustbox}
\end{center}
\caption{Overview of the integral-based key-recovery framework on \cipherone}
\label{fig:framework}
\end{figure}

Based on this idea, we construct a full-cipher distinguisher by extending the integral distinguisher of \branchthree, and use it as the basis for recovering the \branchthree's whitening key bits involved in the Sboxes where there are both constant and active output bits. That is, we proceed an $(r+1)$-round attack based on an $(r+0.5)$-round distinguisher, where the 0.5 round includes all operations except for the Sbox layer. The overall framework is illustrated in Figure~\ref{fig:framework}.

Assume the $(r+0.5)$-round integral distinguisher on \branchthree involves $d$ active input bits. We fix the remaining $n-d$ constant (non-active) bits and enumerate all possible values of the $d$ active bits, thereby forming a set of $2^d$ inputs of $(r+0.5)$-round encryption. Suppose these $n-d$ constant bits cover $N_1$ Sboxes where all the four output bits of the Sbox are constant and $N_2$ Sboxes where the output of Sbox contains both constant and active bits. We guess $2^{4\cdot N_2}$ bits whitening keys involved in these $N_2$ Sboxes and use them to partially decrypt the $(r+0.5)$-round inputs, calculating the corresponding $2^d$ plaintexts. For each guessed key, we can generate $2^d$ ciphertexts by $2^{d}$ queries under the corresponding plaintexts. If the ciphertexts exhibit the balanced property, the guessed key is considered a valid candidate.

% Based on the above framework, we now proceed to mount concrete key-recovery attacks under two different settings: with the full codebook and without the full codebook. In the full-codebook setting, the attacker is allowed to query the cipher on all possible plaintexts, while in the non-full-codebook setting, only a limited number of plaintext queries are allowed.

\subsection{Key-Recovery Attacks on Round-Reduced \cipherone}\label{sect:key_recovery}
We begin by introducing the relevant notations. In \branchthree, the plaintext, the whitening key and the plaintext after XORing with the whitening key are denoted by $X_3=(x_0,x_1,\cdots,$ $ x_{127})\in\mathbb{F}_2^{128}$, $WK=(wk_0,wk_1,\cdots, wk_{127})\in\mathbb{F}_2^{128}$ and $Y=(y_0,y_1,\cdots, y_{127})\in\mathbb{F}_2^{128}$, respectively. Also, $Y$ is the input of Sbox layer in the first round encryption, especially $(y_{4i-4},y_{4i-3},y_{4i-2},y_{4i-1})$ corresponds to the $i^{\text{th}}$ ($1\leq i\leq 32$) Sbox.

\begin{figure}[htp]
\centering
\includegraphics[scale=0.4]{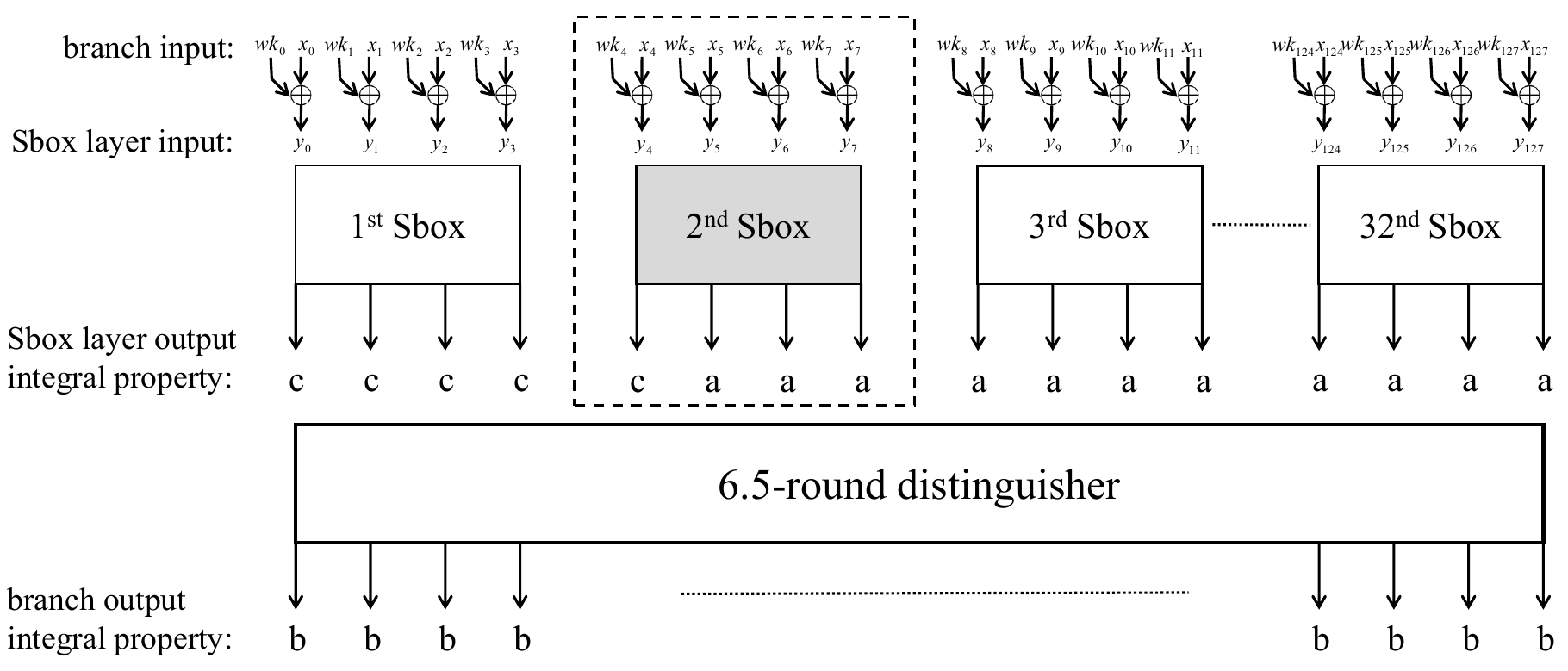}
\caption{Overview of \branchthree in the 7-round attack}
\label{fig:7_round_attack}
\end{figure}

\vspace{5pt}
\noindent\textbf{A 7-round key-recovery attack.} Recall that the 6-round distinguisher of \cipherone has data complexity $2^{123}$, which can be naturally extended to a 6.5-round integral distinguisher as the 0.5 round encryption excludes the nonlinear layer. By prepending the Sbox layer, we can mount a 7-round key-recovery attack. Particularly, in the Sbox layer of the first round encryption of \branchthree, let the 4 out of 5 constant bits cover the output of the first Sbox and the remaining one cover the output of the second Sbox. Moreover, all the remaining 123 output bits of the S-box layer are active, as depicted in Figure~\ref{fig:7_round_attack}. 
Therefore, a set of $2^{123}$ states at the input of Sbox layer can be generated where $(y_0,y_1,y_2,y_3)$ is constant, $(y_4,y_5,y_6,y_7)$ has 8 values and other 120 bits traverse all the $2^{120}$ values. Note the output of the second Sbox corresponding to 8 values of $(y_4,y_5,y_6,y_7)$ has three active bits and one constant bit. We need to guess the four whitening key bits $(wk_4,wk_5,wk_6,wk_7)$, for each of the 16 guessed values, we do:
\begin{enumerate}
    \item Partially decrypt $(y_4,y_5,y_6,y_7)$ to get $(x_4,x_5,x_6,x_7)$ under the guessed key. Then, set $(x_0,x_1,x_2,x_3)$ to a constant and traverse all the $2^{120}$ values\footnote{These 120 active bits on plaintext ensure that the output of 7-round \branchone/\branchtwo always has the balanced integral property as the 7-round degree is upper bounded by $106$.} for $(x_8,x_9,\cdots,x_{127})$ to get $2^{123}$ plaintexts.
    \item Query \cipherone for 7-round ciphertexts under the plaintexts.
    \item Check the integral property of the multiset of ciphertexts. If it is balanced, we regard the guessed value as a candidate.
\end{enumerate}

In fact, the set of plaintexts derived from different key guesses differ only in the values of $(x_4,x_5,x_6,x_7)$, resulting in $2^{124}$ different plaintexts where $(x_0,x_1,x_2,x_3)$ are fixed and the remaining 124 bits span all possible combinations. Moreover, the reason why the correct key in the above process cannot be uniquely determined is that certain wrongly guessed keys cause the second Sbox output to preserve the 3-bit active property after partial encryption of the plaintext. 

To clarify how many candidates of the guessed key remain, we conduct experimental tests on the Sbox. Specifically, we construct four sets of 8 inputs for the Sbox, denoted by $\mathbb{S}_i$ ($0\leq i \leq 3$), where $\mathbb{S}_i$ can ensure that the $i^{\text{th}}$ output bit of the Sbox is constant 0, others are active. When encryption is performed under an incorrect key, it is equivalent to XORing every value in set $\mathbb{S}_i$ with a non-zero constant, where this constant precisely equals the differential value between the correct and wrong keys. Consequently, we apply 16 constants (from 0\textup{x}0 to 0\textup{x}f) to each set $\mathbb{S}_i$ to investigate the active properties of the corresponding Sbox output sets, and record cases exhibiting 3-bit activity as shown in Table~\ref{tab:sbox_active_test}.
\begin{table}[!htbp]
    \centering
    \small
    \caption{The input set of the Sbox and the difference of the correct key and survived wrong key. IP is the integral property of output set of $\mathbb{S}_i$ through the Sbox $S_4$, where `c' and `a' mean constant and active properties, respectively.}
    \label{tab:sbox_active_test}
    \setlength{\tabcolsep}{5pt}
    \begin{tabular}{c|c|c|c|c}
    \toprule
     Input set    &  $\mathbb{S}_0$ & $\mathbb{S}_1$ & $\mathbb{S}_2$ & $\mathbb{S}_3$\\\hline%\midrule
     IP & caaa & acaa & aaca & aaac \\\hline
     Difference    & 0x0, 0x5 & 0x0, 0x1, 0xe & 0x0, 0x8, 0xf & 0x0, 0xe, 0xf\\
     \bottomrule
    \end{tabular}
\end{table}

The results demonstrate that the number of candidates depends on the selection of the 8 inputs to the Sbox, and there are 1-2 wrong keys surviving for different $\mathbb{S}_i$. It is worth noting that the differential intersection of $\mathbb{S}_0$ and $\mathbb{S}_1$ is 0x0, meaning that the intersection of their corresponding candidate key sets yields the correct key. Hence, we can exploit $\mathbb{S}_0$ and $\mathbb{S}_1$ to mount attacks and obtain their respective candidates. Then the correct one can be uniquely determined by taking the intersection of the two candidate sets. The complete attack process to recover $(wk_4,wk_5,wk_6,wk_7)$ is revisited as follows:
\begin{enumerate}
     \item Under $\mathbb{S}_0$, calculate 16 sets of $2^{123}$ plaintexts by partial decrytion for 16 guessed keys and query the 7-round \cipherone to generate 16 sets of $2^{123}$ ciphertexts. Check the integral property for each set of ciphertexts. The guessed keys are retained as candidates if the corresponding sets of ciphertexts have balanced property, forming a set of key candidates $\mathbb{K}_c^0$.
    \item Under $\mathbb{S}_1$, similarly to step 1, a set of candidates $\mathbb{K}_c^1$ can be derived.
    \item The correct value of $(wk_4,wk_5,wk_6,wk_7)$ is determined as $\mathbb{K}_c^0 \cap \mathbb{K}_c^1$.
\end{enumerate}
As analyzed earlier, the data complexity is $2^{124}$ chosen plaintexts. Regardless of the time complexity of recovering $(wk_4,wk_5,wk_6,wk_7)$, it is mainly contributed by the $2^{123}\times 2^{4} \times 2= 2^{128}$ 7-round encryptions. According to the key schedule, these four recovered whitening key bits uniquely correspond to four bits of the 256-bit master key. As a result, to recover the full 256-bit master key, the total time complexity is $2^{128} + 2^{252} \approx 2^{252}$.  

\vspace{5pt}
\noindent\textbf{An improved 7-round key-recovery attack.} The aforementioned 7-round attack recovers only four key bits and remaining 252 key bits are recovered by the brute-force attack. To reduce the time complexity, we aim to extract more key bits information by further exploiting the integral property. Recall the above 7-round attack, we fix the one constant bit on the second Sbox so that we recover $(wk_4,wk_5,wk_6,wk_7)$. In fact, we can relocate this constant bit to other Sboxes e.g, the third one while the four constant bits still stay in the first Sbox. In this case, we can recover the whitening key bits $(wk_8,wk_9,wk_{10},wk_{12})$ with the same $2^{124}$ chosen plaintexts as used in the previous 7-round attack. That is, we can recover the 124 bits of whitening key ($wk_i$ for $4 \leq i \leq 127$) by sliding the single constant bit from the $2^{\text{nd}}$ Sbox to the $32^{\text{nd}}$ Sbox. As a consequence, the time complexity is significantly decreased from $2^{252}$ to 
\[
    2^{123}\times 2^{4}\times 2 \times 31+2^{132}\approx 2^{133.6}
\]    
whereas the data complexity remains unchanged. Moreover, if we pre-compute the $2^{124}$ 7-round plaintext-ciphertext pairs and store them in a table, causing a memory complexity of $2^{124}\times 2^{5} = 2^{129}$ bytes, then the time complexity can be further reduced to
\[
    2^{124} + 2^{132} \approx 2^{132}.
\]

\vspace{5pt}
\noindent\textbf{An 8-round key-recovery attack within the full codebook.} If we use the 7.5-round distinguisher with the data complexity $2^{127}$ to mount the 8-round attack, the full codebook is required as there is only one constant bit in the output of the prepended Sbox layer. In this case, the constant bit can be positioned at any Sbox so that all the 128 whitening key bits can be recovered. As a result, the time complexity is
\[
  2^{127}\times 2^{4} \times 2 \times 32 + 2^{128} = 2^{137}.  
\]
Consider storing the codebook in the pre-processing phase with a memory complexity of $2^{133}$ bytes, the time complexity is reduced to $2^{128} + 2^{128} = 2^{129}$, consisting of querying the full codebook and the exhaustive key search over the remaining 128 bits.

\section{Exploring New Linear-layer Parameters for \branchthree against Linear Attacks}\label{sect:linear_distinguisher}
When searching for DL distinguishers of \cipherone's \branchthree, we find that the linear distinguishers, as reported in the design document~\cite{DBLP:journals/tches/AnandBCIILMRS24}, are false for round numbers exceeding 2. 
% Therefore, we conduct an independent re-evaluation of the linear behavior of \branchthree. Also, the new distinguishers for lower rounds are experimentally verified, supporting the correctness of our revised analysis. The corrected squared correlations of \branchthree of \cipherone are listed in Table~\ref{tab:branch3-linear}. 
Upon further investigation, we identify the reason: the authors incorrectly characterized the linear mask propagation through $\theta$ operation when utilizing SAT-based automatic tool. For \branchthree of \cipherone, the $\theta$ operation can be regarded as a linear transformation over $\mathbb{F}_2^{128}$. Assuming the corresponding transformation matrix is $M$, the input and output are $\bm{x}=(x_{0},x_{1},\cdots,x_{127})^{T}$ and $\bm{y}=(y_{0},y_{1},\cdots,y_{127})^{T}$, respectively, then $\bm{y} = M\cdot \bm{x}$. Denote the mask corresponding to $\bm{x}$ (resp. $\bm{y}$) by $\bm{a}=(a_0,a_1,\cdots,a_{127})^T$ (resp. $\bm{b} = (b_0,b_1,\cdots,b_{127})^T$). According to the mask propagation rules through XOR and branching operations, the relation of input and output masks can be accurately expressed by the following transformation over $\mathbb{F}_2^{128}$:
\[
    \bm{a} = M^{T}\cdot \bm{b}.
\]
However, the design document may mistakenly describe this relation as $\bm{b} = M\cdot \bm{a}$. We re-characterize the mask propagation through $\theta$ using above exact relation and update the linear distinguishers as shown in Table~\ref{tab:branch3-linear}. Also, the new distinguishers for lower rounds are experimentally verified, supporting the correctness of our revised analysis. 
\begin{table}[!htbp]
    \centering
    \small
    \caption{The original~\cite{DBLP:journals/tches/AnandBCIILMRS24} and our corrected squared correlations of \textsf{Branch3}'s linear distinguisher of \cipherone.}
    \label{tab:branch3-linear}
    \renewcommand\tabcolsep{5pt}
    \begin{tabular}{ccccccccc}
    \toprule
     Round & 1 & 2 & 3 & 4 & 5 & 6 & 7 & 8\\\midrule
     \cite{DBLP:journals/tches/AnandBCIILMRS24} & $2^{-2}$ & $2^{-8}$ & $2^{-20}$  & $2^{-30}$ & $2^{-36}$ & $2^{-48}$ & $2^{-52}$ & $2^{-64}$ \\
    Ours& $2^{-2}$ & $2^{-8}$ & $\bm{2^{-12}}$ & $\bm{2^{-16}}$ & $\bm{2^{-20}}$ & $\bm{2^{-24}}$ &$\bm{2^{-28}}$ & $\bm{2^{-32}}$ \\
     \bottomrule
    \end{tabular}
\end{table}

Notably, our evaluation reveal that the $r$-round (for $r>1$) optimal linear distinguisher has squared correlation $2^{-4\cdot r}$. This result indicates that the full-round \branchthree can be distinguished with the expected data complexity $2^{48}$, exposing a significantly weaker resistance to linear attacks and undermining the overall security of the PRF. Therefore, it is of great significance to strengthen the security of \branchthree. Given this vulnerability, it becomes imperative to reinforce the linear cryptanalytic strength of \branchthree. A direct and effective countermeasure is to redesign its linear layer.

To enhance the linear resistance of \branchthree while preserving the original design rationale, we aim to redesign the $\theta$ and $\pi$ layers by tuning their parameters. Our goal is to achieve improved linear security without compromising diffusion properties. Following the original approach, the $\theta$ operation is constructed by XORing three bit and $\pi$ is a bit-permutation. Given $n = 128$ and $gcd(p,128)=1$, there are in total $64 \times \binom{128}{3}$ candidates. Among them, we retain only those ensuring full diffusion within 2.5 rounds, yielding 4352 viable configurations. As in the original design, we further evaluate the minimum number of influenced bits at 0.5 rounds before full diffusion, resulting in 4096 candidates that maintain the same level of diffusion as the original design.

To evaluate their linear resistance, we apply SAT-based automatic tools to compute the maximal squared correlation over the first four rounds. Among the 4096 candidates, 3136 reproduce the same linear profile as the original (i.e., squared correlations: 2, 8, 12, 16). Notably, the remaining 960 parameter sets yield a stronger linear profile, indicating a significantly enhanced resistance against linear cryptanalysis in the early rounds. The newly identified parameter sets are listed in Supplementary Material Table~\ref{tab:NewParams}. To illustrate the enhanced security, we select representative sets from each category and conduct a thorough evaluation of their linear resistance over extended rounds. The results are summarized in Table~\ref{tab:improved-linear}.

\begin{table}[!htbp]
\centering
\small
\caption{Squared correlation of the optimal linear distinguishers up to eight rounds for different parameter categories of \branchthree.}
\label{tab:improved-linear}
\renewcommand{\arraystretch}{1.2}
\setlength{\tabcolsep}{8pt}
\resizebox{\linewidth}{!}{
\parbox{\linewidth}{
\begin{tabular}{lccccccccc}
\toprule
\textbf{Category} & \textbf{Count} & \textbf{1} & \textbf{2} & \textbf{3} & \textbf{4} & \textbf{5}& \textbf{6}& \textbf{7} & \textbf{8}\\
\midrule
Original Design & 3136 & $2^{-2}$ & $2^{-8}$ & $2^{-12}$ & $2^{-16}$ & $2^{-20}$ & $2^{-24}$& $2^{-28}$& $2^{-32}$\\
Improved Class A & 64 & $2^{-2}$ & $2^{-6}$ & $2^{-14}$ & $2^{-24}$ & $2^{-34}/2^{-36}$ & $2^{-46}/2^{-48}$ & $2^{-54}/2^{-58}$ & $2^{-66}$ \\
Improved Class B & 64 & $2^{-2}$ & $2^{-6}$ & $2^{-14}$ & $2^{-26}$ & $2^{-42}$ & $2^{-54}$ & $2^{-66}$ & $2^{-78}$\\
Improved Class C & 256 & $2^{-2}$ & $2^{-8}$ & $2^{-14}$ & $2^{-28}$ & $2^{-46}$ & $2^{-62}$ & $2^{-74}/2^{-76}$ & $2^{-84}$\\
Improved Class D & 64 & $2^{-2}$ & $2^{-8}$ & $2^{-16}$ & $2^{-28}$ & $2^{-36}$ & $2^{-44}$ & $2^{-52}$ & $2^{-60}$\\
Improved Class E & 512 & $2^{-2}$ & $2^{-8}$ & $2^{-16}$ & $2^{-32}$ & $2^{-36}$ & $2^{-48}$ & $2^{-52}$ & $2^{-64}$\\
\bottomrule
\end{tabular}
}}

\vspace{0.5em}
\begin{minipage}{0.95\linewidth}
\small
\textbf{Notes:} We classify all 4096 candidates based on their squared correlation up to round 4. For higher-round evaluations, results are merged within each class.
\end{minipage}
\end{table}
\vspace{-8pt}
From the results in Table~\ref{tab:improved-linear}, it is evident that parameter adjustment improves security while preserving the same diffusion property. We select the well-performing Improved Classes B and C as potential candidates. In addition, differential characteristics are also taken into consideration. Ultimately, we recommend Improved Class B as the optimal choice. This recommendation is justified by its markedly stronger resistance to linear attacks compared with the original design, along with its overall balanced performance among all improved classes, while still maintaining strong differential properties. The detailed differential results are presented in Table~\ref{tab:improve_diff}.

\begin{table}[!htbp]
\centering
\small
\caption{The probability of optimal differential characteristics of \branchthree for the Improved Class B and C.}
\label{tab:improve_diff}
\renewcommand{\arraystretch}{1.2}
\setlength{\tabcolsep}{8pt}
\resizebox{\linewidth}{!}{
\parbox{\linewidth}{
\begin{tabular}{lcccccccc}
\toprule
\textbf{Category} & \textbf{1} & \textbf{2} & \textbf{3} & \textbf{4} & \textbf{5}& \textbf{6}& \textbf{7} & \textbf{8}\\
\midrule
Original Design &  $2^{-2}$ & $2^{-8}$ & $2^{-20}$ & $2^{-32}$ & $2^{-36}$ & $2^{-48}$& $2^{-52}$& $2^{-64}$\\
Improved Class B &  $2^{-2}$ & $2^{-8}$ & $2^{-15}$ & $2^{-27}/2^{-25}$ & $2^{-44}/2^{-45}$ &$\leq 2^{-62}$ & $ -$ &$-$ \\
Improved Class C &  $2^{-2}$ & $2^{-8}$ & $2^{-12}$ & $2^{-16}$ & $2^{-20}$ &$2^{-24}$ & $ 2^{-28}$ &$2^{-32}$ \\
\bottomrule
\end{tabular}
}}

\vspace{0.5em}
\begin{minipage}{0.95\linewidth}
\small
\textbf{Notes:} (1) The notation “$\leq$” denotes that the reported value is a upper bound, since the optimal solution could not be obtained within the given time limit. (2) A dash ($-$) indicates that the result could not be obtained within the given time limit.
\end{minipage}
\end{table}

\section{Conclusion}\label{sec:conclusion}
In this paper, we present the first third-party cryptanalysis of \cipherone, providing a comprehensive evaluation that encompasses DL and integral distinguishers, as well as key-recovery attacks targeting the overall PRF. We propose a two-stage MILP-based method for accurately identifying high-quality DL distinguishers, achieving reduced data complexity compared to previous results. By analyzing algebraic degree bounds, we construct the longest known integral distinguishers for \cipherone and design an attack framework that addresses the inherent challenges posed by its parallel branching structure. Moreover, we reveal weaknesses in \branchthree's linear resistance caused by incorrect mask propagation modeling in the original specification, and propose optimized linear-layer parameters that significantly enhance its security. Furthermore, the proposed frameworks in this work can naturally extend to \ciphertwo. Our results do not threaten the security of \cipherone, but we believe these contributions not only advance the cryptanalysis of \cipherone, but also provide a deeper understanding for designing the multi-branch ciphers.

\backmatter

% \bmhead{Supplementary information}

% If your article has accompanying supplementary file/s please state so here. 

% Authors reporting data from electrophoretic gels and blots should supply the full unprocessed scans for key as part of their Supplementary information. This may be requested by the editorial team/s if it is missing.

% Please refer to Journal-level guidance for any specific requirements.

\bmhead{Acknowledgements}
We would like to thank all the anonymous reviewers for their professional and insightful comments, which help us to significantly improve the quality of this paper. This work was supported by the National Natural Science Foundation of China (Grant No. 62272147, 12471492, 62072161, and 12401687), the Innovation Group Project of the Natural Science Foundation of Hubei Province of China (Grant No. 2023AFA021), the National Key Research and Development Project (Grant No. 2018YFA0704705), Shandong Provincial Natural Science Foundation (Grant No. ZR2024QA205) and the CAS Project for Young Scientists in Basic Research (Grant No. YSBR-035).

% \begin{itemize}
% \item Conflict of interest/Competing interests (check journal-specific guidelines for which heading to use)
% \item Ethics approval and consent to participate
% \item Consent for publication
% \item Materials availability
% \end{itemize}

\begin{appendices}

\section{Truth table, DDT, LAT, and DLCT of \texorpdfstring{$S_3$, $S_4$, and $S_5$}{S3, S4, and S5}}
In this section, we first give the truth tables of $S_3$, $S_4$, and $S_5$, and then present the DDT, LAT, and DLCT, respectively. Based on the entries in these tables, feasible points can be generated to represent valid differential, linear, or DL transitions, which will be further used to construct MILP-based models.

\begin{table*}[!htbp]
\centering
\caption{Truth tables of Sboxes used in \branchone, \branchtwo, and \branchthree. All Sboxes are given in hexadecimal.}
\label{tab:sboxes}

\vspace{1em} % ⬅ 加一点空行，避免主 caption 紧贴 (a)

% ===================== (a) S3 =====================
\begin{center}
(a) Sbox $S_3$
\end{center}
\makebox[\textwidth][c]{%
\begin{tabular}{|>{\centering\arraybackslash}p{2.3em}|
                *{8}{>{\centering\arraybackslash}p{0.8em}|}}
\hline
$x$ & 0 & 1 & 2 & 3 & 4 & 5 & 6 & 7\\ \hline
$S_3(x)$ & 0 & 5 & 3 & 2 & 6 & 1 & 4 & 7\\
\hline
\end{tabular}
}

\vspace{1.5em}

% ===================== (b) S4 =====================
\begin{center}
(b) Sbox $S_4$
\end{center}
\makebox[\textwidth][c]{%
\begin{tabular}{|>{\centering\arraybackslash}p{2.3em}|
                *{16}{>{\centering\arraybackslash}p{0.8em}|}}
\hline
$x$ & 0 & 1 & 2 & 3 & 4 & 5 & 6 & 7 & 8 & 9 & a & b & c & d & e & f\\ \hline
$S_4(x)$ & 1 & 0 & 2 & 4 & 3 & 8 & 6 & d & 9 & a & b & e & f & c & 7 & 5\\
\hline
\end{tabular}
}

\vspace{1.5em}

% ===================== (c) S5 =====================
\begin{center}
(c) Sbox $S_5$
\end{center}
\makebox[\textwidth][c]{%
\begin{tabular}{|>{\centering\arraybackslash}p{2.3em}|
                *{16}{>{\centering\arraybackslash}p{0.8em}|}}
\hline
$x$        & 0 & 1 & 2 & 3 & 4 & 5 & 6 & 7 & 8 & 9 & a & b & c & d & e & f\\ \hline
$S_5(x)$   & 0 & 5 & a & b & 14 & 11 & 16 & 17 & 9 & c & 3 & 2 & d & 8 & f & e\\
\hline\hline
$x$        & 10 & 11 & 12 & 13 & 14 & 15 & 16 & 17 & 18 & 19 & 1a & 1b & 1c & 1d & 1e & 1f\\ \hline
$S_5(x)$   & 12 & 15 & 18 & 1b & 6 & 1 & 4 & 7 & 1a & 1d & 10 & 13 & 1e & 19 & 1c & 1f\\
\hline
\end{tabular}
}
\end{table*}

% ================== Table A1. DDT for S3 ==================
\makebox[\textwidth][c]{%
\begin{minipage}{0.8\textwidth}
\centering
\captionof{table}{DDT for $S_3$}
\label{tab:S3DDT}
\renewcommand{\arraystretch}{1.2}
\scriptsize
\setlength{\tabcolsep}{2pt}

\begin{adjustbox}{max width=\textwidth, max height=0.7\textheight}
\begin{tabular}{c|cccccccc}
\toprule
\diagbox{$\Delta_{in}$}{$\Delta_{out}$} & 0 & 1 & 2 & 3 & 4 & 5 & 6 & 7 \\
\midrule
0 & 8 & 0 & 0 & 0 & 0 & 0 & 0 & 0  \\
1 & 0 & 2 & 0 & 2 & 0 & 2 & 0 & 2  \\
2 & 0 & 0 & 2 & 2 & 0 & 0 & 2 & 2  \\
3 & 0 & 2 & 2 & 0 & 0 & 2 & 2 & 0  \\
4 & 0 & 0 & 0 & 0 & 2 & 2 & 2 & 2  \\
5 & 0 & 2 & 0 & 2 & 2 & 0 & 2 & 0  \\
6 & 0 & 0 & 2 & 2 & 2 & 2 & 0 & 0  \\
7 & 0 & 2 & 2 & 0 & 2 & 0 & 0 & 2  \\
\bottomrule
\end{tabular}
\end{adjustbox}
\end{minipage}
}

\vspace{1em}

% ================== Table A2. DDT for S4 ==================
\makebox[\textwidth][c]{%
\begin{minipage}{0.8\textwidth}
\centering
\captionof{table}{DDT for $S_4$}
\label{tab:S4DDT}
\renewcommand{\arraystretch}{1}
\scriptsize
\setlength{\tabcolsep}{2pt}

\begin{adjustbox}{max width=\textwidth, max height=0.7\textheight}
\begin{tabular}{c|cccccccccccccccc}
\toprule
\diagbox{$\Delta_{in}$}{$\Delta_{out}$}
    & 0 & 1 & 2 & 3 & 4 & 5 & 6 & 7 & 8 & 9 & a & b & c & d & e & f \\
    \midrule
    0 & 16 & 0 & 0 & 0 & 0 & 0 & 0 & 0 & 0 & 0 & 0 & 0 & 0 & 0 & 0 & 0  \\
    1 & 0 & 2 & 2 & 4 & 0 & 2 & 2 & 0 & 0 & 0 & 0 & 4 & 0 & 0 & 0 & 0  \\
    2 & 0 & 0 & 2 & 2 & 4 & 4 & 0 & 0 & 2 & 2 & 0 & 0 & 0 & 0 & 0 & 0 \\
    3 & 0 & 2 & 2 & 0 & 0 & 2 & 0 & 2 & 0 & 0 & 2 & 2 & 0 & 0 & 4 & 0 \\
    4 & 0 & 0 & 2 & 0 & 2 & 0 & 4 & 0 & 2 & 2 & 0 & 2 & 2 & 0 & 0 & 0  \\
    5 & 0 & 0 & 2 & 2 & 0 & 4 & 0 & 0 & 0 & 4 & 0 & 0 & 0 & 0 & 2 & 2  \\
    6 & 0 & 2 & 2 & 0 & 2 & 0 & 0 & 2 & 0 & 0 & 0 & 0 & 2 & 2 & 2 & 2  \\
    7 & 0 & 2 & 0 & 0 & 0 & 0 & 2 & 4 & 0 & 0 & 2 & 0 & 4 & 2 & 0 & 0  \\
    8 & 0 & 2 & 0 & 0 & 2 & 0 & 0 & 0 & 4 & 2 & 4 & 0 & 2 & 0 & 0 & 0  \\
    9 & 0 & 0 & 0 & 2 & 0 & 0 & 0 & 2 & 0 & 2 & 2 & 2 & 2 & 0 & 0 & 4  \\
    a & 0 & 2 & 0 & 0 & 2 & 0 & 0 & 0 & 0 & 2 & 2 & 2 & 0 & 2 & 4 & 0  \\
    b & 0 & 0 & 2 & 0 & 0 & 0 & 2 & 0 & 2 & 0 & 2 & 2 & 0 & 2 & 0 & 4  \\
    c & 0 & 2 & 2 & 2 & 0 & 2 & 0 & 0 & 0 & 0 & 2 & 0 & 2 & 2 & 2 & 0 \\
    d & 0 & 2 & 0 & 2 & 0 & 0 & 2 & 2 & 2 & 2 & 0 & 0 & 0 & 2 & 0 & 2  \\
    e & 0 & 0 & 0 & 0 & 0 & 2 & 4 & 2 & 4 & 0 & 0 & 0 & 0 & 2 & 0 & 2  \\
    f & 0 & 0 & 0 & 2 & 4 & 0 & 0 & 2 & 0 & 0 & 0 & 2 & 2 & 2 & 2 & 0  \\
    \bottomrule
\end{tabular}
\end{adjustbox}
\end{minipage}
}

\vspace{1em}

% ================== Table A3. DDT for S5 ==================
\makebox[\textwidth][c]{%
\begin{minipage}{0.8\textwidth}
\centering
\captionof{table}{DDT for $S_5$}
\label{tab:S5DDT}
\renewcommand{\arraystretch}{1}
\scriptsize
\setlength{\tabcolsep}{2pt}

\begin{adjustbox}{max width=\textwidth, max height=0.7\textheight}
\begin{tabular}{c|*{32}{c}}
\toprule
\diagbox{$\Delta_{in}$}{$\Delta_{out}$}& 0 & 1 & 2 & 3 & 4 & 5 & 6 & 7 & 8 & 9 & a & b & c & d & e & f & 10 & 11 & 12 & 13 & 14 & 15 & 16 & 17 & 18 & 19 & 1a & 1b & 1c & 1d & 1e & 1f \\
\midrule
 0 & 32 & 0 & 0 & 0 & 0 & 0 & 0 & 0 & 0 & 0 & 0 & 0 & 0 & 0 & 0 & 0 & 0 & 0 & 0 & 0 & 0 & 0 & 0 & 0 & 0 & 0 & 0 & 0 & 0 & 0 & 0 & 0 \\
    1 & 0 & 8 & 0 & 8 & 0 & 8 & 0 & 8 & 0 & 0 & 0 & 0 & 0 & 0 & 0 & 0 & 0 & 0 & 0 & 0 & 0 & 0 & 0 & 0 & 0 & 0 & 0 & 0 & 0 & 0 & 0 & 0 \\
    2 & 0 & 0 & 8 & 0 & 0 & 0 & 8 & 0 & 0 & 0 & 8 & 0 & 0 & 0 & 8 & 0 & 0 & 0 & 0 & 0 & 0 & 0 & 0 & 0 & 0 & 0 & 0 & 0 & 0 & 0 & 0 & 0 \\
    3 & 0 & 4 & 0 & 4 & 0 & 4 & 0 & 4 & 0 & 4 & 0 & 4 & 0 & 4 & 0 & 4 & 0 & 0 & 0 & 0 & 0 & 0 & 0 & 0 & 0 & 0 & 0 & 0 & 0 & 0 & 0 & 0 \\
    4 & 0 & 0 & 0 & 0 & 8 & 0 & 0 & 0 & 0 & 0 & 0 & 0 & 8 & 0 & 0 & 0 & 0 & 0 & 0 & 0 & 8 & 0 & 0 & 0 & 0 & 0 & 0 & 0 & 8 & 0 & 0 & 0 \\
    5 & 0 & 4 & 0 & 4 & 0 & 0 & 0 & 0 & 0 & 0 & 0 & 0 & 0 & 4 & 0 & 4 & 0 & 4 & 0 & 4 & 0 & 0 & 0 & 0 & 0 & 0 & 0 & 0 & 0 & 4 & 0 & 4 \\
    6 & 0 & 0 & 4 & 0 & 0 & 0 & 4 & 0 & 0 & 0 & 4 & 0 & 0 & 0 & 4 & 0 & 0 & 0 & 4 & 0 & 0 & 0 & 4 & 0 & 0 & 0 & 4 & 0 & 0 & 0 & 4 & 0 \\
\bottomrule
\end{tabular}
\end{adjustbox}
\end{minipage}
}

%%%%%%%%%%%%%%%%%%%%%%%%%%%%%%%%%%%%%%%%%%%%%%%%%%%%%%
%%%%%%%%%%%%%%%   LAT for S3   %%%%%%%%%%%%%%%%%%%%%%%
%%%%%%%%%%%%%%%%%%%%%%%%%%%%%%%%%%%%%%%%%%%%%%%%%%%%%%
\makebox[\textwidth][c]{%
\begin{minipage}{0.8\textwidth}
\centering
\captionof{table}{LAT for $S_3$}
\label{tab:S3LAT}
\renewcommand{\arraystretch}{1.2}
\scriptsize
\setlength{\tabcolsep}{2pt}
\begin{adjustbox}{max width=\textwidth, max height=0.7\textheight}
\begin{tabular}{c|cccccccc}
\toprule
\diagbox{$\lambda_{in}$}{$\lambda_{out}$} & 0 & 1 & 2 & 3 & 4 & 5 & 6 & 7 \\
\midrule
0 & 4 & 0 & 0 & 0 & 0 & 0 & 0 & 0 \\
1 & 0 & 2 & 0 & 2 & 0 & -2 & 0 & 2 \\
2 & 0 & 0 & 2 & -2 & 0 & 0 & 2 & 2 \\
3 & 0 & 2 & -2 & 0 & 0 & 2 & 2 & 0 \\
4 & 0 & 0 & 0 & 0 & 2 & 2 & -2 & 2 \\
5 & 0 & -2 & 0 & 2 & 2 & 0 & 2 & 0 \\
6 & 0 & 0 & 2 & 2 & -2 & 2 & 0 & 0 \\
7 & 0 & 2 & 2 & 0 & 2 & 0 & 0 & -2 \\
\bottomrule
\end{tabular}
\end{adjustbox}
\end{minipage}
}

%%%%%%%%%%%%%%%%%%%%%%%%%%%%%%%%%%%%%%%%%%%%%%%%%%%%%%
%%%%%%%%%%%%%%%   LAT for S4   %%%%%%%%%%%%%%%%%%%%%%%
%%%%%%%%%%%%%%%%%%%%%%%%%%%%%%%%%%%%%%%%%%%%%%%%%%%%%%
\makebox[\textwidth][c]{%
\begin{minipage}{0.8\textwidth}
\centering
\captionof{table}{LAT for $S_4$}
\label{tab:S4LAT}
\renewcommand{\arraystretch}{1.2}
\scriptsize
\setlength{\tabcolsep}{2pt}
\begin{adjustbox}{max width=\textwidth, max height=0.7\textheight}
\begin{tabular}{c|cccccccccccccccc}
\toprule
\diagbox{$\lambda_{in}$}{$\lambda_{out}$} &
0 & 1 & 2 & 3 & 4 & 5 & 6 & 7 & 8 & 9 & a & b & c & d & e & f \\
\midrule
 0 & 8 & 0 & 0 & 0 & 0 & 0 & 0 & 0 & 0 & 0 & 0 & 0 & 0 & 0 & 0 & 0  \\
    1 & 0 & -4 & -4 & 0 & 2 & -2 & 2 & -2 & 2 & 2 & -2 & -2 & 0 & 0 & 0 & 0  \\
    2 & 0 & 0 & 2 & 2 & 4 & 0 & 2 & -2 & -2 & -2 & 0 & 0 & 2 & -2 & 0 & 4 \\
    3 & 0 & -4 & 2 & -2 & -2 & -2 & 0 & 0 & 0 & 0 & 2 & 2 & 2 & -2 & -4 & 0  \\
    4 & 0 & 2 & 0 & -2 & 4 & -2 & 0 & -2 & 0 & 2 & 4 & 2 & 0 & 2 & 0 & -2  \\
    5 & 0 & -2 & 4 & -2 & 2 & 0 & -2 & 0 & -2 & 0 & -2 & -4 & 0 & 2 & 0 & -2  \\
    6 & 0 & -2 & 2 & -4 & 0 & 2 & 2 & 0 & 2 & 0 & 0 & 2 & -2 & 0 & 4 & 2  \\
    7 & 0 & 2 & 2 & 0 & -2 & -4 & 4 & -2 & 0 & -2 & -2 & 0 & -2 & 0 & 0 & -2  \\
    8 & 0 & 2 & 2 & 0 & 2 & 0 & 0 & 2 & 4 & 2 & -2 & 0 & 2 & -4 & 0 & -2  \\
    9 & 0 & 2 & -2 & -4 & 0 & -2 & -2 & 0 & 2 & -4 & 0 & -2 & 2 & 0 & 0 & 2  \\
    a & 0 & -2 & 0 & 2 & 2 & 0 & 2 & 4 & 2 & -4 & 2 & 0 & 0 & 2 & 0 & -2  \\
    b & 0 & -2 & 0 & 2 & 0 & -2 & -4 & -2 & 0 & -2 & 0 & 2 & 0 & -2 & 4 & -2  \\
    c & 0 & 0 & 2 & 2 & -2 & 2 & 0 & -4 & 4 & 0 & 2 & -2 & 2 & 2 & 0 & 0  \\
    d & 0 & 0 & -2 & -2 & 0 & 4 & 2 & -2 & -2 & -2 & 0 & 0 & 2 & -2 & 0 & -4  \\
    e & 0 & 0 & 0 & 0 & -2 & -2 & 2 & 2 & -2 & 2 & 2 & -2 & 4 & 0 & 4 & 0  \\
    f & 0 & 0 & 0 & 0 & 0 & 0 & 0 & 0 & 0 & 0 & 4 & -4 & -4 & -4 & 0 & 0  \\
\bottomrule
\end{tabular}
\end{adjustbox}
\end{minipage}
}

%%%%%%%%%%%%%%%%%%%%%%%%%%%%%%%%%%%%%%%%%%%%%%%%%%%%%%
%%%%%%%%%%%%%%%   LAT for S5   %%%%%%%%%%%%%%%%%%%%%%%
%%%%%%%%%%%%%%%%%%%%%%%%%%%%%%%%%%%%%%%%%%%%%%%%%%%%%%
\makebox[\textwidth][c]{%
\begin{minipage}{0.8\textwidth}
\centering
\captionof{table}{LAT for $S_5$}
\label{tab:S5LAT}
\renewcommand{\arraystretch}{1.2}
\scriptsize
\setlength{\tabcolsep}{2pt}
\begin{adjustbox}{max width=\textwidth, max height=0.7\textheight}
\begin{tabular}{c|*{32}{c}}
\toprule
\diagbox{$\lambda_{in}$}{$\lambda_{out}$}& 
0 & 1 & 2 & 3 & 4 & 5 & 6 & 7 & 8 & 9 & a & b & c & d & e & f & 
10 & 11 & 12 & 13 & 14 & 15 & 16 & 17 & 18 & 19 & 1a & 1b & 1c & 1d & 1e & 1f \\
\midrule
0 & 16 & 0 & 0 & 0 & 0 & 0 & 0 & 0 & 0 & 0 & 0 & 0 & 0 & 0 & 0 & 0 & 0 & 0 & 0 & 0 & 0 & 0 & 0 & 0 & 0 & 0 & 0 & 0 & 0 & 0 & 0 & 0 \\
    1 & 0 & 8 & 0 & 0 & 0 & 0 & 0 & 0 & 0 & 4 & 0 & 4 & 0 & -4 & 0 & 4 & 0 & -8 & 0 & 0 & 0 & 0 & 0 & 0 & 0 & 4 & 0 & 4 & 0 & -4 & 0 & 4 \\
    2 & 0 & 0 & 8 & -8 & 0 & 0 & 0 & 0 & 0 & 0 & 0 & 0 & 0 & 0 & 0 & 0 & 0 & 0 & 4 & 4 & 0 & 0 & 4 & 4 & 0 & 0 & -4 & -4 & 0 & 0 & 4 & 4 \\
    3 & 0 & 0 & -8 & 0 & 0 & 0 & 0 & 0 & 0 & 4 & 0 & 4 & 0 & 4 & 0 & -4 & 0 & 0 & 4 & -4 & 0 & 0 & 4 & 4 & 0 & 4 & -4 & 0 & 0 & 4 & 4 & 0 \\
    4 & 0 & 0 & 0 & 0 & 8 & 4 & -8 & 4 & 0 & 0 & 0 & 0 & 0 & 4 & 0 & 4 & 0 & 0 & 0 & 0 & 0 & -4 & 0 & -4 & 0 & 0 & 0 & 0 & 0 & 4 & 0 & 4 \\
    5 & 0 & 0 & 0 & 0 & 8 & 4 & 8 & -4 & 0 & -4 & 0 & 4 & 0 & 0 & 0 & 0 & 0 & 0 & 0 & 0 & 0 & -4 & 0 & 4 & 0 & -4 & 0 & 4 & 0 & 0 & 0 & 0 \\
    6 & 0 & 0 & 0 & 0 & -8 & 4 & 0 & -4 & 0 & 0 & 0 & 0 & 0 & 4 & 0 & 4 & 0 & 0 & 4 & 4 & 0 & -4 & 4 & 0 & 0 & 0 & 4 & 4 & 0 & 4 & -4 & 0 \\
\bottomrule
\end{tabular}
\end{adjustbox}
\end{minipage}
}

%%%%%%%%%%%%%%%%%%%%%%%%%%%%%%%%%%%%%%%%%%%%%%%%%%%%%%
%%%%%%%%%%%%%%%   DLCT for S3   %%%%%%%%%%%%%%%%%%%%%%%
%%%%%%%%%%%%%%%%%%%%%%%%%%%%%%%%%%%%%%%%%%%%%%%%%%%%%%
\makebox[\textwidth][c]{%
\begin{minipage}{0.8\textwidth}
\centering
\captionof{table}{DLCT for $S_3$}
\label{tab:S3DLCT}
\renewcommand{\arraystretch}{1.2}
\scriptsize
\setlength{\tabcolsep}{2pt}
\begin{adjustbox}{max width=\textwidth, max height=0.7\textheight}
\begin{tabular}{c|cccccccc}
\toprule
\diagbox{$\Delta_{in}$}{$\lambda_{out}$} & 0 & 1 & 2 & 3 & 4 & 5 & 6 & 7 \\
\midrule
0 & 4 & 4 & 4 & 4 & 4 & 4 & 4 & 4  \\
1 & 4 & -4 & 0 & 0 & 0 & 0 & 0 & 0  \\
2 & 4 & 0 & -4 & 0 & 0 & 0 & 0 & 0  \\
3 & 4 & 0 & 0 & -4 & 0 & 0 & 0 & 0  \\
4 & 4 & 0 & 0 & 0 & -4 & 0 & 0 & 0  \\
5 & 4 & 0 & 0 & 0 & 0 & -4 & 0 & 0  \\
6 & 4 & 0 & 0 & 0 & 0 & 0 & -4 & 0  \\
7 & 4 & 0 & 0 & 0 & 0 & 0 & 0 & -4  \\
\bottomrule
\end{tabular}
\end{adjustbox}
\end{minipage}
}

%%%%%%%%%%%%%%%%%%%%%%%%%%%%%%%%%%%%%%%%%%%%%%%%%%%%%%
%%%%%%%%%%%%%%%   DLCT for S4   %%%%%%%%%%%%%%%%%%%%%%%
%%%%%%%%%%%%%%%%%%%%%%%%%%%%%%%%%%%%%%%%%%%%%%%%%%%%%%
\makebox[\textwidth][c]{%
\begin{minipage}{0.8\textwidth}
\centering
\captionof{table}{DLCT for $S_4$}
\label{tab:S4DLCT}
\renewcommand{\arraystretch}{1.2}
\scriptsize
\setlength{\tabcolsep}{2pt}
\begin{adjustbox}{max width=\textwidth, max height=0.7\textheight}
\begin{tabular}{c|cccccccccccccccc}
\toprule
\diagbox{$\Delta_{in}$}{$\lambda_{out}$} &
0 & 1 & 2 & 3 & 4 & 5 & 6 & 7 & 8 & 9 & a & b & c & d & e & f \\
\midrule
 0 & 8 & 8 & 8 & 8 & 8 & 8 & 8 & 8 & 8 & 8 & 8 & 8 & 8 & 8 & 8 & 8  \\
    1 & 8 & -4 & -4 & 0 & 4 & -4 & -4 & 4 & 4 & 0 & 0 & -4 & 0 & 0 & 0 & 0  \\
    2 & 8 & 0 & 4 & 0 & 0 & 0 & -4 & 0 & 4 & 0 & 0 & 0 & -4 & 0 & -8 & 0  \\
    3 & 8 & 0 & -4 & -4 & 0 & 0 & 0 & 0 & 0 & -4 & 4 & 0 & 0 & 4 & 0 & -4  \\
    4 & 8 & 4 & 0 & 0 & 0 & -4 & 0 & 0 & 0 & 4 & -4 & -4 & -4 & 0 & 0 & 0  \\
    5 & 8 & -4 & 0 & -4 & 0 & 0 & 0 & 0 & 0 & 0 & 0 & 0 & 0 & 4 & -8 & 4  \\
    6 & 8 & 0 & 0 & 0 & -4 & 0 & 0 & -4 & 0 & 0 & 0 & 0 & 4 & 0 & 0 & -4  \\
    7 & 8 & 0 & 0 & 0 & -4 & 0 & 0 & -4 & 0 & -4 & -4 & 0 & 0 & 0 & 8 & 0  \\
    8 & 8 & 4 & 4 & 0 & 4 & 0 & 0 & -4 & -4 & -4 & 0 & 0 & -4 & -4 & 0 & 0  \\
    9 & 8 & -4 & -4 & 4 & 0 & 0 & 0 & -4 & -4 & 0 & 0 & 0 & 0 & 0 & 0 & 4  \\
    a & 8 & 0 & 0 & -4 & 0 & -4 & 0 & 0 & -4 & 0 & 4 & 4 & 0 & 0 & 0 & -4  \\
    b & 8 & 0 & -4 & 0 & 0 & 4 & 0 & 0 & -4 & 4 & 0 & -4 & 0 & -4 & 0 & 0  \\
    c & 8 & 0 & 0 & -4 & 0 & 0 & -4 & 0 & 0 & -4 & 0 & 0 & 4 & 0 & 0 & 0  \\
    d & 8 & -4 & 0 & 0 & 0 & 0 & 4 & 0 & 0 & 0 & -4 & 0 & 0 & -4 & 0 & 0  \\
    e & 8 & 0 & 0 & 0 & -4 & 4 & 4 & 4 & 0 & 0 & -4 & -4 & -4 & -4 & 0 & 0  \\
    f & 8 & 0 & 0 & 4 & -4 & -4 & -4 & 0 & 0 & 0 & 0 & 4 & 0 & 0 & 0 & -4  \\
\bottomrule
\end{tabular}
\end{adjustbox}
\end{minipage}
}

%%%%%%%%%%%%%%%%%%%%%%%%%%%%%%%%%%%%%%%%%%%%%%%%%%%%%%
%%%%%%%%%%%%%%%   DLCT for S5   %%%%%%%%%%%%%%%%%%%%%%%
%%%%%%%%%%%%%%%%%%%%%%%%%%%%%%%%%%%%%%%%%%%%%%%%%%%%%%
\makebox[\textwidth][c]{%
\begin{minipage}{0.8\textwidth}
\centering
\captionof{table}{DLCT for $S_5$}
\label{tab:S5DLCT}
\renewcommand{\arraystretch}{1.2}
\scriptsize
\setlength{\tabcolsep}{2pt}
\begin{adjustbox}{max width=\textwidth, max height=0.7\textheight}
\begin{tabular}{c|*{32}{c}}
\toprule
\diagbox{$\Delta_{in}$}{$\lambda_{out}$}& 
0 & 1 & 2 & 3 & 4 & 5 & 6 & 7 & 8 & 9 & a & b & c & d & e & f & 
10 & 11 & 12 & 13 & 14 & 15 & 16 & 17 & 18 & 19 & 1a & 1b & 1c & 1d & 1e & 1f \\
\midrule
 0 & 16 & 16 & 16 & 16 & 16 & 16 & 16 & 16 & 16 & 16 & 16 & 16 & 16 & 16 & 16 & 16 & 16 & 16 & 16 & 16 & 16 & 16 & 16 & 16 & 16 & 16 & 16 & 16 & 16 & 16 & 16 & 16\\
    1 & 16 & -16 & 0 & 0 & 0 & 0 & 0 & 0 & 16 & -16 & 0 & 0 & 0 & 0 & 0 & 0 & 16 & -16 & 0 & 0 & 0 & 0 & 0 & 0 & 16 & -16 & 0 & 0 & 0 & 0 & 0 & 0\\
    2 & 16 & 16 & -16 & -16 & 0 & 0 & 0 & 0 & 0 & 0 & 0 & 0 & 0 & 0 & 0 & 0 & 16 & 16 & -16 & -16 & 0 & 0 & 0 & 0 & 0 & 0 & 0 & 0 & 0 & 0 & 0 & 0\\
    3 & 16 & -16 & 0 & 0 & 0 & 0 & 0 & 0 & 0 & 0 & 0 & 0 & 0 & 0 & 0 & 0 & 16 & -16 & 0 & 0 & 0 & 0 & 0 & 0 & 0 & 0 & 0 & 0 & 0 & 0 & 0 & 0\\
    4 & 16 & 16 & 16 & 16 & -16 & -16 & -16 & -16 & 0 & 0 & 0 & 0 & 0 & 0 & 0 & 0 & 0 & 0 & 0 & 0 & 0 & 0 & 0 & 0 & 0 & 0 & 0 & 0 & 0 & 0 & 0 & 0\\
    5 & 16 & -16 & 0 & 0 & 0 & 0 & 0 & 0 & 0 & 0 & 0 & 0 & 16 & -16 & 0 & 0 & 0 & 0 & 0 & 0 & 0 & 0 & 0 & 0 & 0 & 0 & 0 & 0 & 0 & 0 & 0 & 0\\
    6 & 16 & 16 & -16 & -16 & 0 & 0 & 0 & 0 & 0 & 0 & 0 & 0 & 0 & 0 & 0 & 0 & 0 & 0 & 0 & 0 & 0 & 0 & 0 & 0 & 0 & 0 & 0 & 0 & 0 & 0 & 0 & 0\\
\bottomrule
\end{tabular}
\end{adjustbox}
\end{minipage}
}

\section{Inequalities Used in MILP Models}
\label{Supplementary Material:Ineq}
The inequalities presented in this Supplementary Material are used to model the DDT, LAT and DLCT of Sboxes within the MILP model. These inequalities were obtained by applying the algorithm in \cite{Feng2022FullLI}, following the MILP-based modelling method of Sasaki and Todo~\cite{DBLP:conf/secitc/SasakiT17}, which also enables the selection of a minimal  subset of inequalities.

\begin{itemize}
    \item Encoding the DDT of $S_3$ with $(x_0,x_1,x_2,y_0,y_1,y_2,p)$, where $(x_0,x_1,x_2) \xrightarrow{S_3}(y_0,y_1,y_2)$ denotes the valid differential transition with probability $2^{-2p}$. The DDT is characterized as:
    
\begingroup
\scriptsize
\setlength{\abovedisplayskip}{2pt}
\setlength{\belowdisplayskip}{2pt}
\setlength{\abovedisplayshortskip}{1pt}
\setlength{\belowdisplayshortskip}{1pt}
\allowdisplaybreaks
\begin{equation}
\label{eq:DDTS3}
\begin{cases}
-3x_2-3x_1 + x_0-3y_2-3y_1 + y_0 + 10p \geq 0 \\
-3x_2 + x_1-3x_0-3y_2 + y_1-3y_0 + 10p \geq 0 \\
x_2-3x_1-3x_0 + y_2-3y_1-3y_0 + 10p \geq 0 \\
4x_2 + 7x_1 + 2x_0 + 4y_2 + y_1 + 6y_0-8p \geq 0 \\
4x_2 + 7x_1 + 6x_0 + 4y_2 + y_1 + 2y_0-8p \geq 0 \\
6x_2 + x_1 + 4x_0 + 2y_2 + 7y_1 + 4y_0-8p \geq 0 \\
2x_2 + x_1 + 4x_0 + 6y_2 + 7y_1 + 4y_0-8p \geq 0 \\
\end{cases}
\end{equation}
\endgroup
\item Encoding the LAT of $S_3$ with $(x_0,x_1,x_2,y_0,y_1,y_2,c)$, where $(x_0,x_1,x_2) \xrightarrow{S_3}(y_0,y_1,y_2)$ denotes the valid linear approximation with correlation $2^{-c}$. The DDT is characterized as:

\begingroup
\scriptsize
\setlength{\abovedisplayskip}{2pt}
\setlength{\belowdisplayskip}{2pt}
\setlength{\abovedisplayshortskip}{1pt}
\setlength{\belowdisplayshortskip}{1pt}
\allowdisplaybreaks
\begin{equation}
\label{eq:LATS3}
\begin{cases}
-x_2 + 3x_1-x_0 + 6y_2 + 3y_1 + 6y_0-4c \geq 0 \\
6x_2 + 3x_1-x_0-y_2 + 3y_1 + 6y_0-4c \geq 0 \\
4x_2 + x_1 + 8x_0 + 4y_2 + 7y_1-2y_0-6c \geq 0 \\
3x_2 + 6x_1 + 6x_0 + 3y_2-y_1-y_0-4c \geq 0 \\
-3x_2-3x_1 + x_0-3y_2-3y_1 + y_0 + 10c \geq 0 \\
x_2-3x_1-3x_0 + y_2-3y_1-3y_0 + 10c \geq 0 \\
-3x_2 + x_1-3x_0-3y_2 + y_1-3y_0 + 10c \geq 0 \\
\end{cases}
\end{equation}
\endgroup

\item Encoding the unsigned DLCT of $S_3$ with $(x_0,x_1,x_2,y_0,y_1,y_2)$, where $(x_0,x_1,x_2) \xrightarrow{S_3}(y_0,y_1,y_2)$ denotes the valid DL approximation with constant correlation 1. The unsigned DLCT is characterized as:

\begingroup
\scriptsize
\setlength{\abovedisplayskip}{2pt}
\setlength{\belowdisplayskip}{2pt}
\setlength{\abovedisplayshortskip}{1pt}
\setlength{\belowdisplayshortskip}{1pt}
\allowdisplaybreaks
\begin{equation}
\label{eq:DLCTS3}
\begin{cases}
-2x_2 + 4x_1-6x_0-y_2-7y_1 + 4y_0 \geq -8 \\
4x_2-6x_1-2x_0-7y_2 + 4y_1-y_0 \geq -8 \\
-6x_2-2x_1 + 4x_0 + 4y_2-y_1-7y_0 \geq -8 \\
\end{cases}
\end{equation}
\endgroup

\item Encoding the signed DLCT of $S_3$ with $(x_0,x_1,x_2,y_0,y_1,y_2,s)$, where $s=0$ indicates a positive correlation and $s=1$ indicates a negative correlation. The signed DLCT is characterized as:

\begingroup
\scriptsize
\setlength{\abovedisplayskip}{2pt}
\setlength{\belowdisplayskip}{2pt}
\setlength{\abovedisplayshortskip}{1pt}
\setlength{\belowdisplayshortskip}{1pt}
\allowdisplaybreaks
\begin{equation}
\label{eq:DLCTS3_sgn}
\begin{cases}
-8x_2-3x_1-3x_0 + 5y_2-14y_1 + 2y_0 + 7s \geq -14 \\
-2x_2-8x_1-2x_0-y_2 + 6y_1-11y_0 + 6s \geq -12 \\
-4x_2-4x_1-4x_0-11y_2 + 2y_1-y_0 + 10s \geq -12 \\
-5x_2 + 9x_1-3x_0 + 4y_2-9y_1 + 2y_0-7s \geq -9 \\
5x_2-8x_1 + 3x_0-4y_2 + 9y_1-2y_0-9s \geq -8 \\
\end{cases}
\end{equation}
\endgroup

\item Encoding the DDT of $S_4$ with $(x_0,x_1,x_2,x_3,y_0,y_1,y_2,y_3,p_0,p_1)$, where $(x_0,x_1,x_2,x_3) \xrightarrow{S_4}(y_0,y_1,y_2,y_3)$ denotes the valid differential transition with probability $2^{-(2p_0+p_1)}$. This DDT is characterized as:

\begingroup
\scriptsize
\setlength{\abovedisplayskip}{2pt}
\setlength{\belowdisplayskip}{2pt}
\setlength{\abovedisplayshortskip}{1pt}
\setlength{\belowdisplayshortskip}{1pt}
\allowdisplaybreaks
\begin{equation}
\label{eq:DDTS4}
\begin{cases}
-17x_3-29x_2 + 7x_1-19x_0-8y_3 + 11y_2-24y_1 + 5y_0 + 52p_0 + 85p_1 \geq 0 \\
-3x_3-4x_2-3x_1 + 6x_0-y_3 + 4y_2-12y_0-30p_0-22p_1 \geq -41 \\
-11x_3-25x_2-x_1-3x_0-14y_3-6y_2-7y_1-9y_0 + 51p_0 + 75p_1 \geq 0 \\
-8x_3 + 24x_2 + 14x_1 + 38x_0 + 15y_3-17y_2-39y_1 + y_0 + 5p_0-5p_1 \geq -30 \\
10x_3 + 21x_2-2x_1-x_0-20y_3-22y_2-10y_1 + 2y_0 + 16p_0-6p_1 \geq -39 \\
12x_3-9x_2-12x_1-4x_0 + 3y_3 + 8y_2-12y_1-9y_0 + 2p_0-2p_1 \geq -36 \\
18x_3 + 15x_2 + 18x_1-3x_0-17y_3 + 3y_2-y_1-y_0 + 9p_0-9p_1 \geq -13 \\
6x_3 + 3x_2 + 17x_1 + 14x_0 + 3y_3-14y_2-11y_1-10y_0 + 2p_0-5p_1 \geq -20 \\
4x_3 + 5x_2 + 6x_0 + 2y_3-5y_2 + y_1-2y_0-20p_0-26p_1 \geq -27 \\
11x_3-2x_2 + 4x_1-23x_0 + 11y_3-y_2 + 22y_1 + 21y_0 + 4p_0-4p_1 \geq -7 \\
26x_3 + x_2-6x_1 + 22x_0-12y_3 + 13y_2-8y_1 + 4y_0-85p_0-61p_1 \geq -79 \\
-37x_3 + 17x_2-3x_1-4x_0 + 42y_3-2y_2-4y_1-21y_0-28p_0 + 11p_1 \geq -57 \\
-6x_3-2x_2-6x_1 + 8x_0 + 4y_3-4y_2-8y_1 + 10y_0 + p_0-9p_1 \geq -25 \\
3x_3-11x_2-2x_1-8x_0 + 11y_3-11y_2 + 8y_1-3y_0 + 2p_0-7p_1 \geq -31 \\
13x_3 + 12x_2 + 6x_1-x_0-16y_3-9y_2 + 12y_1-3y_0-57p_0-52p_1 \geq -65 \\
-7x_3-12x_2-16x_1-4x_0-9y_3 + 16y_2 + 16y_1-4y_0 + 2p_0-6p_1 \geq -42 \\
8x_3-27x_2 + 8x_1 + 19x_0 + 49y_3 + 51y_2 + 14y_1 + 5y_0-42p_0 + 10p_1 \geq -4 \\
-8x_3 + 3x_2 + 2x_1 + 2x_0 + 12y_3 + 16y_2 + 13y_1 + 16y_0-4p_0-8p_1 \geq 0 \\
-20x_3-20x_2-8x_1 + 6x_0 + 14y_3-2y_2 + 10y_1 + 20y_0 + 3p_0-11p_1 \geq -41 \\
-27x_3 + 54x_2-2x_1-27x_0 + 23y_3-2y_2 + 24y_1 + 7y_0-15p_0 + 15p_1 \geq -19 \\
\end{cases}
\end{equation}
\endgroup

\item Encoding the LAT of $S_4$ with $(x_0,x_1,x_2,x_3,y_0,y_1,y_2,y_3,c_0,c_1)$, where $(x_0,x_1,x_2,x_3) \xrightarrow{S_4}(y_0,y_1,y_2,y_3)$ denotes the valid linear approximation with correlation $2^{-(c_0+c_1)}$. This LAT is characterized as:

\begingroup
\scriptsize
\setlength{\abovedisplayskip}{2pt}
\setlength{\belowdisplayskip}{2pt}
\setlength{\abovedisplayshortskip}{1pt}
\setlength{\belowdisplayshortskip}{1pt}
\allowdisplaybreaks
\begin{equation}
\label{eq:LATS4}
\begin{cases}
14x_3-3x_2 + 14x_1-7x_0 + 2y_3-11y_2 + y_1-7y_0 + 14c_0 + 12c_1 \geq 0 \\
-3x_3 + x_1 + x_0 + 8y_3 + 9y_2 + 9y_1 + 9y_0-3c_0-3c_1 \geq 0 \\
-9x_3 + x_2-10x_1-6x_0-3y_3-3y_2 + y_1 + 10y_0 + 15c_0-8c_1 \geq -15 \\
20x_3-x_2 + 20x_1 + 15x_0 + 7y_3 + 5y_2-3y_1 + 20y_0-3c_0-13c_1 \geq 0 \\
-4x_3-3x_2-4x_0-9y_3-5y_2-9y_1 + 6y_0 + 32c_0-7c_1 \geq 0 \\
-47x_3 + x_2 + 4x_1-47x_0 + 7y_3 + 5y_2 + 8y_1 + 11y_0 + 77c_0 + 10c_1 \geq 0 \\
2x_3 + 2x_2-6x_1-6x_0 + y_3-6y_2-7y_1 + 4y_0 + 12c_0-5c_1 \geq -12 \\
-8x_3 + 13x_2-3x_1 + 15x_0-10y_3 + 20y_2-28y_1-4y_0 + 25c_0 + 24c_1 \geq 0 \\
-x_3 + 8x_2 + 19x_1 + 6x_0-4y_3-16y_2-12y_1 + 13y_0 + 8c_0 + 15c_1 \geq 0 \\
4x_3-x_2-3x_1-6x_0 + 4y_3-5y_2-11y_1-11y_0 + 13c_0 + 11c_1 \geq -13 \\
-6x_3-2x_2-5x_1-x_0 + 10y_3 + 8y_2 + 3y_0 + 4c_0 + 5c_1 \geq 0 \\
17x_3-x_2 + 9x_1-3x_0-23y_3-7y_2 + 16y_1 + 5y_0 + 8c_0 + 21c_1 \geq 0 \\
-2x_3 + 11x_2-3x_1-7x_0-2y_3-5y_2 + 11y_1 + 3y_0 + 8c_0 + 9c_1 \geq 0 \\
13x_3 + 13x_2-x_1 + 8x_0 + 13y_3 + 5y_2 + 13y_1-3y_0-3c_0-6c_1 \geq 0 \\
-x_3 + 4x_2-x_1 + 4x_0 + 4y_3-4y_2 + 3y_1-4y_0 + 3c_0-c_1 \geq -4 \\
6x_3-8x_1 + 2x_0-4y_3 + 8y_2-y_1 + 4y_0 + 5c_0 + 7c_1 \geq 0 \\
-11x_3-12x_2-9x_1-12x_0 + 6y_3-y_2 + 3y_1-y_0 + 20c_0-6c_1 \geq -20 \\
2x_3-5x_2-5x_1 + 2x_0-5y_3-3y_2 + 5y_1-6y_0 + 10c_0-c_1 \geq -10 \\
-x_3 + 4x_2 + 4x_1-2x_0-4y_3-3y_2 + 4y_1-4y_0 + 6c_0-2c_1 \geq -6 \\
-x_3 + 3x_2 + 3x_1-x_0 + 3y_3 + 2y_2-3y_1-3y_0 + 4c_0-3c_1 \geq -4 \\
2x_3 + 2x_2-18x_1-22x_0-20y_3 + 5y_2 + 5y_1 + 3y_0 + 51c_0 + 4c_1 \geq 0 \\
2x_3-9x_2-9x_1 + 5x_0 + 7y_3-y_2 + 4y_1 + 4y_0 + 9c_0 + 7c_1 \geq -1 \\
-x_3 + 4x_2-x_1 + 4x_0-4y_3 + 4y_2-3y_1-4y_0 + 5c_0-c_1 \geq -5 \\
x_3-3x_2 + x_1-3x_0-3y_3 + 3y_2-2y_1-3y_0 + 7c_0-3c_1 \geq -7 \\
-7x_3-7x_2 + 2x_1 + 3x_0-5y_3-8y_2-7y_1-y_0 + 16c_0-4c_1 \geq -16 \\
26x_3 + 26x_2 + 23x_1 + 17x_0-9y_3-9y_2-3y_1-y_0-c_0-3c_1 \geq 0 \\
-3x_3-16x_2 + 6x_1 + 10x_0-2y_3 + 10y_2 + 11y_1-22y_0 + 21c_0 + 19c_1 \geq 0 \\
x_3-8x_2-8x_1 + x_0 + 3y_3 + 2y_2-8y_1-8y_0 + 37c_0-8c_1 \geq 0 \\
\end{cases}
\end{equation}
\endgroup

\item Encoding the unsigned DLCT of $S_4$ with $(x_0,x_1,x_2,x_3,y_0,y_1,y_2,y_3,c)$, where $(x_0,x_1,x_2,x_3) \xrightarrow{S_4}(y_0,y_1,y_2,y_3)$ denotes the valid DL approximation with correlation $2^{-c}$. This unsigned DLCT is characterized as:

\begingroup
\scriptsize
\setlength{\abovedisplayskip}{2pt}
\setlength{\belowdisplayskip}{2pt}
\setlength{\abovedisplayshortskip}{1pt}
\setlength{\belowdisplayshortskip}{1pt}
\allowdisplaybreaks
\begin{equation}
\label{eq:DLCTS4}
\begin{cases}
31x_3-24x_2 + x_1-4x_0-4y_3-6y_2-10y_1-6y_0-10c \geq -32 \\
17x_3 + 12x_2-9x_1-10x_0-11y_3-3y_2-y_1 + 2y_0-28c \geq -32 \\
-50x_3 + 12x_2 + 6x_1 + 10x_0 + 6y_3 + y_2 + 3y_1-7y_0-c \geq -45 \\
-50x_3-6x_2-x_1-3x_0-2y_3 + 48y_2 + 43y_1 + 47y_0 + 49c \geq -12 \\
23x_3 + 17x_2-50x_1 + 20x_0 + 25y_3 + 4y_2-5y_1-44y_0 + 6c \geq -50 \\
19x_3 + 10x_2 + 50x_1-7x_0-28y_3-21y_2-8y_1-6y_0 + 3c \geq -35 \\
13x_3 + 17x_2-3x_1-30x_0 + 4y_3-14y_2-5y_1 + 20y_0-15c \geq -34 \\
47x_3-42x_2-6x_1-17x_0 + 12y_3-4y_2 + 14y_1-50y_0-11c \geq -65 \\
42x_3-32x_2 + 3x_1 + 12x_0-20y_3-43y_2 + 7y_1-4y_0-5c \geq -52 \\
-9x_3-4x_2 + 11x_1 + 8x_0 + 11y_3 + 2y_2-12y_1 + y_0 + 5c \geq -12 \\
-50x_3 + 8x_2-11x_1-9x_0-10y_3 + 3y_2 + 3y_1 + 4y_0 + 2c \geq -67 \\
31x_3-10x_2 + 4x_1 + 4x_0-26y_3 + 18y_2-16y_1-20y_0 + c \geq -36 \\
42x_3-7x_2-16x_1-4x_0-32y_3 + 6y_2 + 13y_1-12y_0-47c \geq -59 \\
37x_3-13x_2-28x_1-5x_0 + 8y_3-23y_2-19y_1 + 21y_0-4c \geq -46 \\
31x_3-33x_2-9x_1 + 7x_0-2y_3-4y_2-28y_1-8y_0 + 9c \geq -44 \\
19x_3-x_2 + 6x_1-24x_0 + 3y_3-5y_2 + 6y_1 + 3y_0-20c \geq -25 \\
17x_3-7x_2-3x_1-15x_0 + 18y_3-5y_2 + 8y_1-20y_0 + c \geq -25 \\
43x_3-40x_2 + 10x_1-4x_0-6y_3 + 7y_2-50y_1 + 5y_0 + 2c \geq -50 \\
22x_3 + 2x_2-24x_1-5x_0 + 6y_3 + 2y_2-6y_1 + 7y_0-23c \geq -29 \\
-50x_3-46x_2-48x_1 + x_0 + 48y_3 + 49y_2 + 47y_1 + 2y_0 + 49c \geq -94 \\
-49x_3 + 49x_2 + 4x_1 + 2x_0-y_3-48y_2 + 47y_1-50y_0 + 46c \geq -98 \\
\end{cases}
\end{equation}
\endgroup

\item Encoding the signed DLCT of $S_4$ with $(x_0,x_1,x_2,x_3,y_0,y_1,y_2,y_3,c,s)$, where $(x_0,x_1,x_2,x_3) \xrightarrow{S_4}(y_0,y_1,y_2,y_3)$ denotes the valid DL approximation with signed correlation $(-1)^{s}\cdot 2^{-c}$. This signed DLCT is characterized as:

\begingroup
\scriptsize
\setlength{\abovedisplayskip}{2pt}
\setlength{\belowdisplayskip}{2pt}
\setlength{\abovedisplayshortskip}{1pt}
\setlength{\belowdisplayshortskip}{1pt}
\allowdisplaybreaks
\begin{equation}
\label{eq:DLCTS4_sgn}
\begin{cases}
-12x_3-7x_2 + 8x_1 + 3x_0-5y_3-y_2 + 3y_1-13y_0 + 6c + 12s \geq -19 \\
-4x_3 + x_2-24x_1-2x_0 + 20y_3 + 23y_2 + 22y_1-8y_0-8c-8s \geq -30 \\
-4x_3-18x_2 + 2x_1 + x_0 + 10y_3 + 24y_2 + 15y_1 + 29y_0-33c + 7s \geq -22 \\
4x_3 + 4x_2 + 20x_1-19x_0-8y_3 + 6y_2-20y_1 + 10y_0-c + 12s \geq -28 \\
4x_3-2x_2-2x_1-9x_0-15y_3-19y_2-9y_1 + 17y_0-18c-s \geq -56 \\
8x_3 + 6x_2 + 4x_1 + 18x_0-9y_2-3y_1-7y_0-18c + 10s \geq -19 \\
-5x_3-5x_2 + 27x_1-7x_0 + 15y_3-9y_2 + 20y_1 + 18y_0-c-17s \geq -17 \\
-7x_3 + 2x_2 + 5x_1-3x_0 + 5y_3-7y_2 + 7y_1-y_0 + 4c-3s \geq -10 \\
16x_3 + 3x_2 + 4x_1-3x_0-5y_3 + 10y_2 + 11y_1 + 12y_0 + 3c-16s \geq -5 \\
-10x_3 + 20x_2 + 16x_1 + 12x_0-18y_3 + 6y_2-30y_1 + 8y_0-c-19s \geq -48 \\
5x_3-2x_2-2x_1-x_0-3y_3-6y_2 + 5y_1-3y_0 + 3c-4s \geq -12 \\
-18x_3-6x_2 + 2x_1-8x_0 + 2y_3-32y_2 + 10y_1 + 27c + 3s \geq -32 \\
-4x_3-5x_2-4x_1 + x_0 + 21y_3 + 21y_2-13y_1 + 2y_0 + 45c-38s \geq -13 \\
-4x_3 + 13x_2-16x_1-7x_0 + 6y_3-8y_2-20y_1-2y_0 + 5c + 10s \geq -32 \\
2x_3 + x_2 + x_1 + 9x_0 + 7y_3 + 13y_2 + 6y_1-13y_0 + 31c-40s \geq -13 \\
-7x_3-3x_2 + 8x_1-4x_0-6y_3 + 5y_2-6y_1-10y_0 + 6c-2s \geq -22 \\
12x_3-2x_2-12x_1 + 9x_0-2y_3 + 12y_2-3y_1-4y_0 + c-4s \geq -14 \\
3x_3 + x_2 + 2x_1 + y_3-2y_2-3y_1-3y_0 + 2c-5s \geq -8 \\
-44x_3-26x_2 + x_1 + 4x_0 + 4y_3-10y_2-6y_1-54y_0 + 97c-36s \geq -70 \\
15x_3-17x_2-4x_1 + 5x_0 + 9y_3-4y_2-17y_1 + 17y_0-c-2s \geq -28 \\
-2x_3 + 14x_2-25x_1-10x_0 + 2y_3-21y_2 + 3y_1 + 27y_0-6c + 13s \geq -37 \\
-4x_3-27x_2 + 19x_1 + 14x_0 + y_3 + 14y_2-29y_1-6y_0-2c + 35s \geq -35 \\
-6x_3 + 22x_2-23x_1 + 24x_0-8y_3-20y_2 + 10y_1-4y_0-c + 22s \geq -32 \\
-28x_3-13x_2-4x_1-4x_0-49y_3 + 14y_2 + 7y_1 + 5y_0 + 37c + s \geq -49 \\
7x_3 + x_2 + 2x_1-39x_0 + 3y_3 + 8y_2 + 5y_1-39y_0 + 27c + 12s \geq -39 \\
\end{cases}
\end{equation}
\endgroup

\item Encoding the DDT of $S_5$ with $(x_0,x_1,x_2,x_3,x_4,y_0,y_1,y_2,y_3,y_4,p_0,p_1,p_2)$, where $(x_0,x_1,x_2,x_3,x_4) \xrightarrow{S_5}(y_0,y_1,y_2,y_3,y_4)$ denotes the valid differential transition with probability $2^{-(2p_0+3p_1+4p_2)}$. The DDT is characterized as:

\begingroup
\scriptsize
\setlength{\abovedisplayskip}{2pt}
\setlength{\belowdisplayskip}{2pt}
\setlength{\abovedisplayshortskip}{1pt}
\setlength{\belowdisplayshortskip}{1pt}
\allowdisplaybreaks
\begin{equation}
\label{eq:DDTS5}
\begin{cases}
-39x_4-50x_3-21x_2 + 5x_1-23x_0-y_4-9y_3-9y_2-2y_1 + 2y_0-156p_1-112p_2 \geq -260 \\
67x_4-42x_3 + 42x_2-38x_1 + 20x_0 + 68y_4-68y_3 + y_2-10y_1 + 10y_0 + 2p_0-24p_1 + 2p_2 \geq -114 \\
-14x_4-7x_2 + x_0-2y_4 + y_3-7y_2 + 2y_0-33p_0-21p_1-19p_2 \geq -49 \\
x_4-11x_3-15x_2-12x_1-7x_0 + y_4 + y_3 + 2y_2-2y_1-2y_0 + 4p_0 + 26p_1 + 37p_2 \geq -10 \\
-x_4-2x_3-x_2-4x_1-3x_0-3y_4-3y_3 + 3y_2-3y_1-3y_0 + 2p_2 \geq -18 \\
-24x_4-36x_3-39x_2-36x_1-43x_0-13y_4 + y_3 + 4y_2-11y_1 + 6y_0-213p_0-123p_1-60p_2 \geq -261 \\
-33x_4-28x_3-35x_2-45x_1-52x_0-16y_4-5y_3 + 7y_2 + y_1 + 3y_0-225p_0-129p_1-61p_2 \geq -274 \\
-40x_4-41x_3-17x_2 + x_1-17x_0 + 2y_4-3y_3-9y_2 + 8y_1 + 2y_0-138p_1-98p_2 \geq -222 \\
94x_4-8x_3-3x_2-35x_1-10x_0-y_4 + 2y_3-3y_2 + 95y_1 + 95y_0 + 7p_0 + 25p_1-35p_2 \geq 0 \\
-23x_4-19x_3-38x_2-42x_1-34x_0 + 7y_4-y_3 + 2y_2 + 2y_1-7y_0-175p_1-114p_2 \geq -277 \\
-x_4-4x_3-3x_2-x_1-2x_0 + 3y_4-3y_3-3y_2-3y_1-3y_0 + 2p_2 \geq -18 \\
-143x_4-119x_3-120x_2-116x_1-146x_0 + 8y_4-21y_3 + 17y_2-3y_1 + 9y_0 + 55p_0 + 323p_1 + 580p_2 \geq -85 \\
92x_4 + 231x_3-2x_2 + 30x_1-183x_0 + 99y_4-40y_3 + 41y_2 + 18y_1 + 199y_0 + p_0-6p_1 + 4p_2 \geq 0 \\
-25x_4 + x_3-19x_2-25x_1 + 4x_0 + 6y_4 + 5y_3-25y_2-25y_1 + 4y_0-23p_0-4p_1 + 13p_2 \geq -98 \\
-3x_4 + 20x_3-3x_2-5x_1 + 20x_0-4y_4-2y_3 + y_2-y_1-40p_0-50p_2 \geq -48 \\
-2x_4 + 17x_3-64x_2 + 17x_1 + 26x_0 + 3y_4 + 60y_2 + 6y_1-5y_0-202p_0-161p_1-175p_2 \geq -206 \\
-90x_4-115x_3-103x_2 + x_1-58x_0-y_4 + 24y_3 + 21y_2 + 5y_1-5y_0-581p_0-422p_1-308p_2 \geq -678 \\
-40x_4 + 46x_3 + 97x_2-54x_1 + 60x_0-14y_4 + 12y_3 + 98y_2-98y_1 + y_0 + 8p_0-36p_1 + 2p_2 \geq -144 \\
68x_4 + 112x_3-47x_2 + 63x_1-44x_0-3y_4 + 112y_3-112y_2-y_1-5y_0 + 24p_0-38p_1 + 11p_2 \geq -138 \\
28x_4-x_3 + 24x_2-35x_1-31x_0-58y_4 + 57y_3-2y_1-3y_0-51p_0-17p_1 + 17p_2 \geq -88 \\
-58x_4 + 4x_3-69x_2-126x_1-139x_0-17y_4 + 8y_3-7y_2-5y_1-11y_0-102p_1 + 33p_2 \geq -390 \\
-16x_4-16x_3 + 5x_2-8x_1 + x_0-16y_4-16y_3-8y_2-y_1-22p_0-5p_1 + 5p_2 \geq -70 \\
-24x_4-54x_3-60x_2-29x_1 + 5x_0-y_4-3y_3-7y_2-8y_1-98p_0-44p_1 + 12p_2 \geq -169 \\
-4x_4-6x_3-x_2 + x_1 + 2y_4 + 4y_3-15p_0-10p_1 \geq -17 \\
-4x_3-6x_2 + x_0 + 2y_3 + 4y_2-15p_0-10p_1 \geq -17 \\
-20x_4-24x_3-12x_2 + x_1-15x_0 + 4y_3 + 2y_2-48p_0-19p_1 + 2p_2 \geq -68 \\
-8x_4-8x_3 + 26x_2 + 13x_1-x_0 + 13y_4 + 26y_3-y_2 + 2p_0-3p_1-8p_2 \geq 0 \\
134x_4-2x_3 + 67x_2 + 67x_1-4x_0-4y_4-6y_3-11y_1 + 7y_0 + 12p_0 + 20p_1-107p_2 \geq 0 \\
44x_4 + 44x_3-x_2 + 66x_1-4x_0-y_4-4y_3 + 3y_2-5y_1 + 22y_0 + 3p_0-73p_2 \geq 0 \\
13x_4 + 18x_3 + 24x_2 + 9x_1 + 26x_0 + 4y_4-y_2 + y_1-14y_0-3p_0-37p_2 \geq 0 \\
-219x_4 + 32x_3 + 62x_2 + 2x_1 + 21x_0 + 225y_4 + 34y_3 + 3y_2 + 4y_1 + 2y_0-117p_0 + 43p_1 + 21p_2 \geq -111 \\
59x_4 + 4x_3 + 17x_2-209x_1 + 31x_0 + y_4 + 4y_3 + 2y_2 + 217y_1 + 34y_0-114p_0 + 43p_1 + 21p_2 \geq -106 \\
3x_4 + x_3-209x_2-130x_1 + 28x_0 + 2y_4 + y_3 + 130y_2 + 51y_1 + 5y_0-199p_0 + 10p_1 + 28p_2 \geq -278 \\
-47x_4-43x_3-43x_1-35x_0 + 2y_4 + y_3-10y_2 + 4y_1-4y_0-111p_0 + 21p_2 \geq -160 \\
-34x_4 + x_3-34x_2-41x_1-41x_0-8y_4-4y_3-y_2 + y_1-5y_0-100p_0-20p_1 + 19p_2 \geq -147 \\
x_4-8x_3-15x_2-16x_1-11x_0-y_1-2y_0 + 4p_0 + 24p_1 + 39p_2 \geq -13 \\
2x_4-28x_3 + 9x_2 + 3x_1 + 9x_0-36y_2 + 5y_1-23p_0 + 5p_1 + 2p_2 \geq -51 \\
-35x_4-64x_3 + x_2 + 2x_1 + 13y_4 + 37y_3 + 2y_2-5y_1-58p_0 + 2p_1 + 16p_2 \geq -85 \\
-70x_4 + x_3-126x_2-112x_1 + x_0-42y_4 + 28y_2-28y_1 + 3y_0-232p_0 + 3p_1 + 5p_2 \geq -372 \\
-31x_3-29x_2-14x_1-y_4 + 4y_3 + 2y_2-14y_1-53p_0 + 8p_2 \geq -81 \\
-53x_4 + 3x_3-9x_2 + x_1-54x_0-5y_4-59y_3 + 59y_2 + 9y_1-8y_0-84p_0-29p_1 + 32p_2 \geq -153 \\
x_4-28x_3-15x_2 + 15x_1-26x_0-2y_3 + 4y_2 + 30y_1-30y_0-18p_0-7p_1 + 22p_2 \geq -78 \\
-66x_4-29x_3 + 3x_2-30x_1-61x_0-5y_4-7y_3 + 3y_2-y_0-107p_0-46p_1 + 17p_2 \geq -179 \\
12x_4 + 2x_3-74x_2-75x_1-60x_0-2y_4-y_3 + 4y_2 + 8y_1-8y_0-109p_0-38p_1 + 38p_2 \geq -182 \\
-37x_4 + 10x_3-45x_2-84x_1-91x_0-10y_4-2y_3 + 2y_2 + y_1-6y_0-149p_0-63p_1 + 23p_2 \geq -246 \\
13x_4 + 16x_3 + 18x_2 + 16x_1 + 18x_0 + 3y_4-6y_2 + 2y_1-4y_0-8p_0-24p_1-24p_2 \geq 0 \\
-3x_4 + 20x_3 + 77x_2 + 47x_1 + 68x_0 + 23y_4 + 3y_3-62y_2 + 14y_1-7y_0 + p_0-12p_1-19p_2 \geq 0 \\
\end{cases}
\end{equation}
\endgroup

\item Encoding the LAT of $S_5$ with $(x_0,x_1,x_2,x_3,x_4,y_0,y_1,y_2,y_3,y_4,c_0,c_1)$, where $(x_0,x_1,x_2,x_3,x_4) \xrightarrow{S_5}(y_0,y_1,y_2,y_3,y_4)$ denotes the valid linear approximation with correlation $2^{-(c_0+c_1)}$. The LAT is characterized as:

\begingroup
\scriptsize
\setlength{\abovedisplayskip}{2pt}
\setlength{\belowdisplayskip}{2pt}
\setlength{\abovedisplayshortskip}{1pt}
\setlength{\belowdisplayshortskip}{1pt}
\allowdisplaybreaks
\begin{equation}
\label{eq:LATS5}
\begin{cases}
-x_4-3x_3-3x_2-x_1 + 3x_0 + 2y_4-3y_3-3y_2-4y_1-3y_0 + 6c_0 + 6c_1 \geq -6 \\
x_4-3x_3 + x_2 + 5x_1-3x_0 + 2y_2-4y_1-3y_0 + 5c_0 + c_1 \geq -4 \\
-x_4 + 3x_3-x_2-3x_1-3x_0-3y_4-3y_3 + 2y_2-4y_1-3y_0 + 6c_0 + 6c_1 \geq -6 \\
5x_4 + 2x_3-46x_2-5x_1-2x_0 + 23y_4 + 30y_3 + 41y_2-12y_1 + y_0 + 19c_0 + 23c_1 \geq 0 \\
10x_4 + x_3 + 2x_1-5y_4 + y_3 + 8y_1 + 10y_0-5c_1 \geq 0 \\
-x_4-5x_3-x_1 + 9y_4 + 10y_3-y_2-y_1 + 10y_0 + 4c_0-5c_1 \geq 0 \\
-3x_4 + x_3-3x_2 + x_1 + 5x_0-3y_4 + 2y_1-4y_0 + 5c_0 + c_1 \geq -4 \\
-3x_4 + x_3 + 3x_2 + x_1-3x_0-3y_4-3y_3-3y_2 + 2y_1-3y_0 + 5c_0 + 5c_1 \geq -5 \\
x_4 + 2x_2 + 10x_0 + y_4 + 8y_2 + 10y_1-5y_0-5c_1 \geq 0 \\
-4x_3 + 2x_2-4x_0 + 4y_4-y_3 + 2y_2-4y_1-4y_0 + 8c_0-3c_1 \geq -8 \\
3x_4 + x_3-3x_2-3x_1 + x_0-3y_4 + 2y_3-3y_2-3y_1-3y_0 + 5c_0 + 5c_1 \geq -5 \\
-9x_4-x_3-9x_2 + 9x_1-x_0-9y_4 + 8y_3-3y_2-3y_1-9y_0 + 19c_0-3c_1 \geq -19 \\
2x_3 + 10x_1 + x_0 + 8y_3 + 10y_2-5y_1 + y_0-5c_1 \geq 0 \\
10x_3 + x_2 + 2x_0 + 10y_4-5y_3 + y_2 + 8y_0-5c_1 \geq 0 \\
-2x_4 + 7x_3 + 82x_2-4x_1 + 24x_0 + 82y_4 + 58y_3-17y_2 + 17y_1-6y_0-4c_0-49c_1 \geq 0 \\
-20x_4-2x_3 + 2x_1 + x_0 + 18y_4-5y_3 + 11y_1 + 14y_0 + 7c_0 + 9c_1 \geq 0 \\
-x_4-x_2-5x_1-y_4 + 10y_3 + 9y_2 + 10y_1-y_0 + 4c_0-5c_1 \geq 0 \\
2x_4-4x_2-4x_0 + 2y_4-4y_3-4y_2 + 4y_1-3y_0 + 8c_0-c_1 \geq -8 \\
4x_4 + x_3 + 34x_2 + 2x_1-9x_0 + 15y_4 + 24y_3-92y_2 + 2y_1-58y_0 + 67c_0 + 73c_1 \geq 0 \\
x_4 + 25x_3-9x_2 + 2x_1-12x_0 + 3y_4-76y_3-30y_2-46y_1-58y_0 + 102c_0 + 125c_1 \geq 0 \\
-13x_4-2x_3 + 22x_2-13x_1-x_0-47y_4 + 3y_3-54y_2-16y_1-25y_0 + 75c_0 + 92c_1 \geq 0 \\
-8x_4-8x_3 + x_2 + 3x_1 + x_0-8y_4-8y_3-8y_2-8y_1 + 2y_0 + 53c_0-8c_1 \geq 0 \\
-18x_4 + 8x_3 + x_2-18x_1 + x_0-5y_4-5y_3-18y_2-18y_1 + 7y_0 + 92c_0-18c_1 \geq 0 \\
16x_4-29x_3-4x_2-29x_1-6x_0-29y_4-29y_3-11y_2-9y_1 + 12y_0 + 139c_0-9c_1 \geq 0 \\
-x_3-x_1-5x_0-y_4-y_3 + 10y_2 + 9y_1 + 10y_0 + 4c_0-5c_1 \geq 0 \\
3x_4 + x_3 + 3x_2 + 2x_1 + 2x_0-23y_4 + y_3-24y_2 + y_1-24y_0 + 57c_0 + 11c_1 \geq 0 \\
x_4-18x_3-6x_2 + 4x_1 + 10x_0-22y_4-19y_3-23y_2-31y_1-25y_0 + 60c_0 + 83c_1 \geq 0 \\
3x_4 + 6x_3 + 4x_2-14x_1-6x_0-59y_4-41y_3-54y_2-38y_1-40y_0 + 105c_0 + 144c_1 \geq 0 \\
x_4-18x_3 + x_2-18x_1 + 8x_0-18y_4-18y_3 + 7y_2-5y_1-5y_0 + 92c_0-18c_1 \geq 0 \\
\end{cases}
\end{equation}
\endgroup

\item Encoding the unsigned DLCT of $S_5$ with $(x_0,x_1,x_2,x_3,x_4,y_0,y_1,y_2,y_3,y_4)$, where $(x_0,x_1,x_2,x_3,x_4) \xrightarrow{S_5}(y_0,y_1,y_2,y_3,y_4)$ denotes the valid DL approximation with constant correlation 1. The unsigned DLCT is characterized as:

\begingroup
\scriptsize
\setlength{\abovedisplayskip}{2pt}
\setlength{\belowdisplayskip}{2pt}
\setlength{\abovedisplayshortskip}{1pt}
\setlength{\belowdisplayshortskip}{1pt}
\allowdisplaybreaks
\begin{equation}
\label{eq:DLCTS5}
\begin{cases}
7x_4 + x_3 + 11x_2-17x_1-25x_0 + y_4 + 11y_3-39y_2 + 7y_1 + y_0 \geq -42 \\
7x_4-35x_3-28x_2 + 15x_1 + x_0-63y_4 + 17y_3-3y_2 + y_1 + 22y_0 \geq -66 \\
-9x_4-2x_3 + 5x_1-2y_4 + y_3 + 5y_2-9y_1 \geq -11 \\
3x_3 + x_1-8x_0-2y_4 + y_3-8y_1 + 3y_0 \geq -10 \\
x_4-26x_2-19x_1 + 10x_0 + 17y_4-43y_3 + 12y_2-2y_1 + 3y_0 \geq -47 \\
-8x_4 + 3x_2 + x_0 + 3y_4-2y_3 + y_2-8y_0 \geq -10 \\
21x_4-2x_3-7x_2-34x_1-6y_4-2y_3-32y_2 + 21y_1-3y_0 \geq -44 \\
-x_4-59x_3-7x_2-7x_1 + 37x_0-3y_4-8y_3-10y_2 + 37y_1-57y_0 \geq -78 \\
2x_4 + 30x_3-22x_2-22x_1-6x_0-3y_4-41y_3 + 18y_2-8y_1 \geq -52 \\
\end{cases}
\end{equation}
\endgroup

\item Encoding the signed DLCT of $S_5$ with $(x_0,x_1,x_2,x_3,x_4,y_0,y_1,y_2,y_3,y_4,s)$, where $s=0$ indicates a positive correlation and $s=1$ indicates a negative correlation. The signed DLCT is characterized as:

\begingroup
\scriptsize
\setlength{\abovedisplayskip}{2pt}
\setlength{\belowdisplayskip}{2pt}
\setlength{\abovedisplayshortskip}{1pt}
\setlength{\belowdisplayshortskip}{1pt}
\allowdisplaybreaks
\begin{equation}
\label{eq:DLCTS5_sgn}
\begin{cases}
21x_4 + 42x_3 + 34x_2 + x_1 + 5x_0 + 21y_4-8y_3-32y_2 + y_1 + 5y_0-42s \geq -40 \\
-10x_4 + x_3-42x_1-17x_0 + 13y_4 + 11y_3-52y_2 + 37y_1-8y_0-2s \geq -69 \\
-14x_3 + x_2-5x_1-12y_4 + 14y_3-4y_2 + 5y_1-2s \geq -19 \\
-2x_4-20x_2 + 18x_0 + y_4-14y_3-7y_2-4y_0 + 7s \geq -25 \\
3x_3-3x_0-2y_1-y_0 + s \geq -3 \\
-3x_4 + 3x_2-y_4-2y_0 + s \geq -3 \\
2x_4 + 4x_3-84x_2 + x_1 + 51x_0 + 2y_4-92y_3 + 51y_2 + 9y_1-14y_0-18s \geq -106 \\
-x_4-6x_3 + 5x_1-4y_4-2y_3-y_1 + 2s \geq -7 \\
21x_4-36x_2 + x_1 + 2x_0-40y_4 + 4y_2 + 7y_1 + 21y_0-8s \geq -40 \\
x_4-3x_3 + 27x_2 + 5x_1-111x_0 + 3y_4 + 27y_3-47y_2-69y_1 + 71y_0-12s \geq -119 \\
x_4 + 23x_3-39x_1 + 4x_0 + 23y_4-41y_3 + 6y_1 + 4y_0-8s \geq -41 \\
-33x_4-x_3 + 2x_2 + 19x_1 + 3y_4 + 7y_3 + 19y_2-36y_1-7s \geq -37 \\
-36x_3 + x_2 + 2x_1 + 21x_0 + 4y_3 + 7y_2 + 21y_1-40y_0-8s \geq -40 \\
3x_4-3x_1-2y_2-y_1 + s \geq -3 \\

\end{cases}
\end{equation}
\endgroup
\end{itemize}

To exclude a specific feasible solution $(x_0, x_1, \ldots, x_{m-1}) = (\delta_0, \delta_1, \ldots, \delta_{m-1})$ from the solution space, we add the following \textit{cutting-off inequality} to the MILP model:

\begin{equation}\label{equ:cutting_off}
\sum_{i=0}^{m-1} (-1)^{\delta_i} x_i + \sum_{i=0}^{m-1} \delta_i \ge 1.
\end{equation}

This inequality ensures that the current solution is removed from the feasible region while keeping all other valid solutions, thus allowing the MILP solver to discover new solutions in subsequent iterations.

\section{Newly Identified Linear-layer Parameter Sets}

\renewcommand{\arraystretch}{1.1}
\setlength{\tabcolsep}{6pt}
{\centering
\begingroup\small
\emergencystretch=2em
\begin{longtable}{@{} l >{\tiny\arraybackslash}p{0.78\linewidth} @{}}
\caption{New linear-layer parameters grouped by category.}\label{tab:NewParams}\\
\toprule
\textbf{Category} & \textbf{\normalsize Parameters ($p, t_0,t_1,t_2$)} \\
\midrule
\endfirsthead

\toprule
\textbf{Category} & \textbf{\normalsize Parameters ($p, t_0,t_1,t_2$)}  \\
\midrule
\endhead

\midrule \multicolumn{2}{r}{Continued on next page}\\
\endfoot

\bottomrule
\endlastfoot
Improved Class A & (35,\penalty 200 0,\penalty 200 42,\penalty 200 85); (35,\penalty 200 1,\penalty 200 44,\penalty 200 86); (35,\penalty 200 2,\penalty 200 45,\penalty 200 88); (35,\penalty 200 4,\penalty 200 46,\penalty 200 89); (35,\penalty 200 5,\penalty 200 48,\penalty 200 90); (35,\penalty 200 6,\penalty 200 49,\penalty 200 92); (35,\penalty 200 8,\penalty 200 50,\penalty 200 93); (35,\penalty 200 9,\penalty 200 52,\penalty 200 94); (35,\penalty 200 10,\penalty 200 53,\penalty 200 96); (35,\penalty 200 12,\penalty 200 54,\penalty 200 97); (35,\penalty 200 13,\penalty 200 56,\penalty 200 98); (35,\penalty 200 14,\penalty 200 57,\penalty 200 100); (35,\penalty 200 16,\penalty 200 58,\penalty 200 101); (35,\penalty 200 17,\penalty 200 60,\penalty 200 102); (35,\penalty 200 18,\penalty 200 61,\penalty 200 104); (35,\penalty 200 20,\penalty 200 62,\penalty 200 105) \\
  & (35,\penalty 200 21,\penalty 200 64,\penalty 200 106); (35,\penalty 200 22,\penalty 200 65,\penalty 200 108); (35,\penalty 200 24,\penalty 200 66,\penalty 200 109); (35,\penalty 200 25,\penalty 200 68,\penalty 200 110); (35,\penalty 200 26,\penalty 200 69,\penalty 200 112); (35,\penalty 200 28,\penalty 200 70,\penalty 200 113); (35,\penalty 200 29,\penalty 200 72,\penalty 200 114); (35,\penalty 200 30,\penalty 200 73,\penalty 200 116); (35,\penalty 200 32,\penalty 200 74,\penalty 200 117); (35,\penalty 200 33,\penalty 200 76,\penalty 200 118); (35,\penalty 200 34,\penalty 200 77,\penalty 200 120); (35,\penalty 200 36,\penalty 200 78,\penalty 200 121); (35,\penalty 200 37,\penalty 200 80,\penalty 200 122); (35,\penalty 200 38,\penalty 200 81,\penalty 200 124); (35,\penalty 200 40,\penalty 200 82,\penalty 200 125); (35,\penalty 200 41,\penalty 200 84,\penalty 200 126) \\
\end{longtable}
\endgroup
\par}
\end{appendices}


\bibliography{sn-bibliography}% common bib file

\end{document}